# Learning joint probabilistic weather forecasts from station observations alone

**Chaeyeon Yi[1,3], Yun Am Seo[2,3,*]**

[1] Research Center for Atmospheric Environment, Hankuk University of Foreign Studies, Yongin 17035, Republic of Korea

[2] Department of Data Science, Jeju National University, Jeju 63243, Republic of Korea

[3] NAVI Hyper-Tropicalization Research Institute, Jeju National University, Jeju 63243, Republic of Korea

∗ Corresponding author: Yun Am Seo (seoya@jejunu.ac.kr)

ORCID — Chaeyeon Yi: 0000-0003-2802-2431 · Yun Am Seo: 0000-0001-9283-4376

## Abstract

Assessing compound weather risks requires forecasts representing dependence between variables. CLARA (Calibrated Advection-Routing Attention) learns joint Gaussian predictive distributions of five surface variables from station observations alone, without numerical weather prediction or reanalysis; the approximately 28,000-parameter model supports CPU training and prediction. Across six multi-year folds on 96 stations, its lead-mean energy score is 4.9% lower than that of a learned comparator with matched temporal inputs (4.7% with a similar parameter count) and 11–65% lower than those of statistical baselines. Holding marginal variances fixed, removing learned correlations worsens joint negative log-likelihood by 1.0–2.8 nats per station. A covariance-scale estimator, proved consistent under stated assumptions, improves short-lead calibration but over-corrects at long leads. Synthetic interventions show an attention-bias coefficient alone does not measure forecast influence. Retrained in ten regions on six continents, CLARA outperforms persistence in all 60 multi-year region–lead comparisons and a similarly sized learned model in 57 of 60.

Short-range surface forecasts usually rely on numerical weather prediction (NWP), followed by statistical post-processing against observations to correct errors and uncertainties (Vannitsem et al., 2021). Recent machine-learning systems forecast directly from observations: end-to-end systems ingest heterogeneous observations and issue gridded and station forecasts (Allen et al., 2025), and a global system is trained and initialized from observations alone (Pinnington et al., 2026). Limits of gridded NWP for station-level surface prediction have also been reported (Wu et al., 2023). These systems are, however, mostly deterministic.

Derived quantities and compound risks depend on both marginal accuracy and cross-variable dependence. An explicit joint predictive distribution makes that dependence available for prediction and diagnostic comparison. Station networks add a second question: how the contribution of neighbouring observations changes with lead. At short leads an upwind observation is informative downwind; as the lead lengthens, spatial information decays while the target-hour diurnal cycle and the season gain in importance.

Four gaps motivate this model. First, multivariate post-processing requires NWP forecasts or ensembles (Gneiting et al., 2005; Schefzik et al., 2013), while observation-centred systems often retain reanalysis-pretrained modules or assimilated fields (Allen et al., 2025; Pinnington et al., 2026). Second, fitting probabilistic models separately by variable, station or lead limits information sharing (Rasp & Lerch, 2018; Wessel et al., 2024). Third, marginals do not determine the dependence needed for derived quantities and compound events (Schefzik et al., 2013). Fourth, wind-informed spatial models rarely condition the strength of an upwind bias on forecast lead (Supplementary Note 6).

We propose CLARA (Calibrated Advection-Routing Attention), a single model that addresses these gaps, uses no NWP or reanalysis at any stage, and predicts six leads spanning 1–48 h from station observations at base time. To our knowledge no study integrates these capabilities in one observation-based system (Extended Data Table 1).

The study establishes three results. First, CLARA jointly learns five-variable means and full covariances from station observations, and a lead-wise scale estimator, proved consistent under stated assumptions, recalibrates the covariance while preserving correlations; the explicit covariance makes the value of the learned dependence measurable, and the model supports CPU training and prediction. Second, the strength of its advection-routing bias is a learnable function of lead, $\gamma(\tau)$. Because attention weighs that strength against the content-path scale (Lemma 1), such a coefficient cannot be read on its own; we measure both and test this interpretation on a synthetic task in which the dependence is known. Third, evaluation on two nationwide networks, two held-out seasons and ten regions on six continents quantifies the scope of these results.

## Results

### The model in brief

CLARA outputs a five-variable Gaussian predictive distribution for each station and lead directly from the same base-time observations, without recursive error propagation (Fig. 1). Spatial information is combined by attention over neighbouring stations, biased towards the upwind position of station $i$, the point reached from its position $p_i$ by moving against the base-time wind $w_i$ for a time $\tau$ along the sphere, with a strength $\gamma(\tau)$ learned as a function of lead.

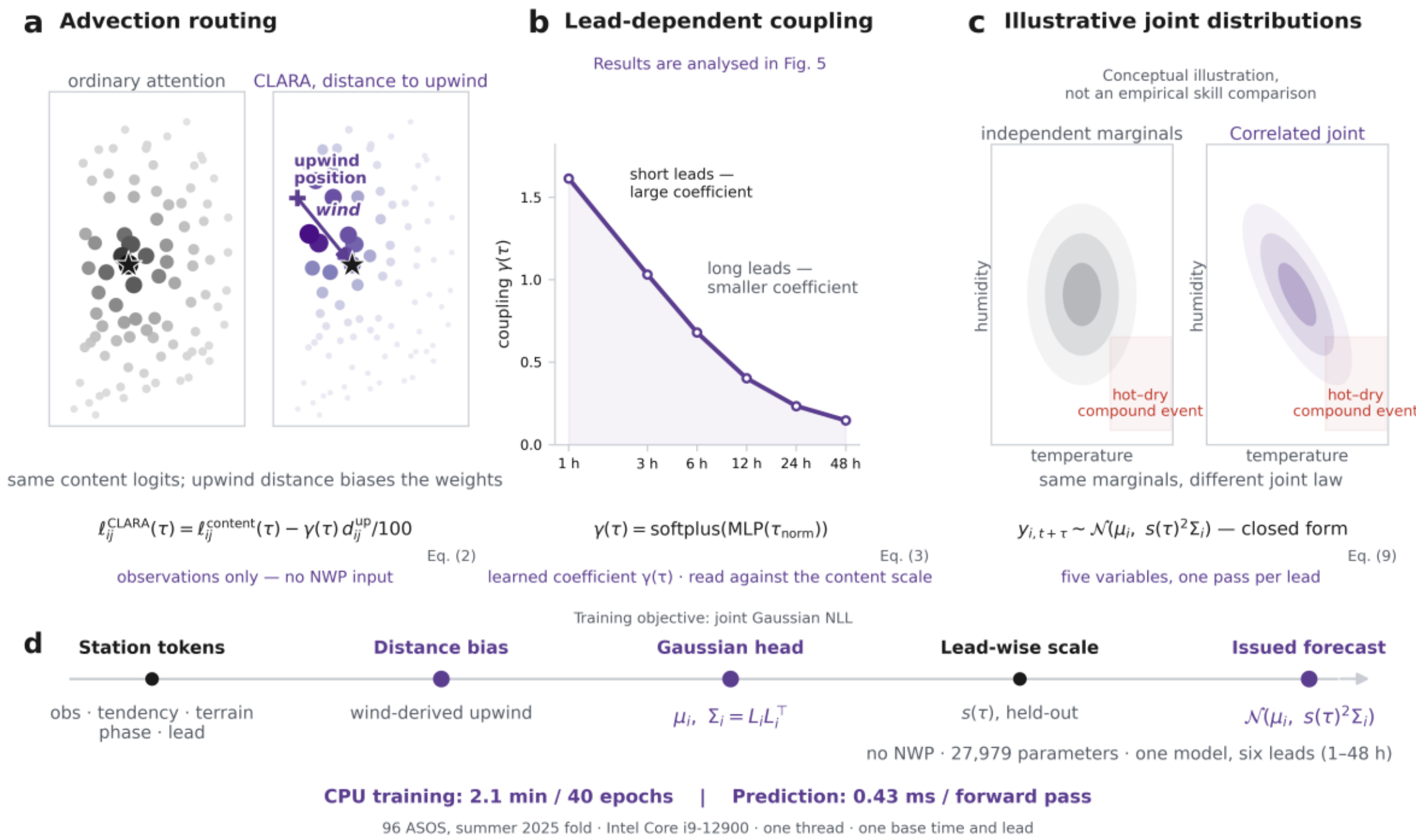

*Figure 1. CLARA produces lead-conditioned joint predictive distributions directly from station observations.* ***a****, Advection routing: for the same stations and target (star), both maps use the same illustrative content logits; CLARA subtracts a term proportional to the distance from a wind-derived upwind location, shifting weight towards it (illustrative weights, displayed relative to their within-map maximum; great-circle distance).* ***b****, The learned coupling γ(τ) (96 ASOS, fold mean of seed means over the six multi-year folds) decreases with lead; its effect must be read against the content scale (Lemma 1); results are analysed in Fig. 5.* ***c****, Conceptual illustration, not an empirical skill comparison: with the same marginals, a joint law with correlation assigns a different probability to a compound hot–dry event than independent marginals.* ***d****, Prediction pipeline from station tokens, including temporal phase and lead, to the issued forecast, one forward pass per lead. The joint Gaussian NLL is the training objective. CPU timings below d cover full training and a forward pass excluding input assembly (96 ASOS, summer 2025; Supplementary Table 9).*

The mean is $\mu_i = y_i + \delta_i$, the base-time observation plus a predicted increment, and the covariance is $\Sigma_i = L_i L_i^\top$ from a Cholesky factor; one Gaussian negative log-likelihood penalizes the mean error and the scale and correlation of the uncertainty together. After training, only the scale is recalibrated, $\Sigma_i' = s(\tau)^2 \Sigma_i$, which preserves the correlation matrix; Proposition 1 gives the assumptions under which $s(\tau)$ consistently estimates the misspecification scale. With 27,979 parameters, CLARA produces the joint distribution for one base time and lead at the 96 stations in a 0.43-ms forward pass on one CPU thread; training for 40 epochs takes 2.1 min (Fig. 1d; Methods).

## Overall forecast accuracy

On the 96 stations of South Korea's Automated Synoptic Observing System (ASOS), the issued joint distribution of CLARA has a lower energy score than every comparison model at all six evaluated leads (fold mean; Fig. 2, Extended Data Table 2), and a lower score in 143 of the 144 fold–lead cells against the learned model with matched temporal-phase input and the three statistical baselines (persistence, climatology and ensemble model output statistics, EMOS).

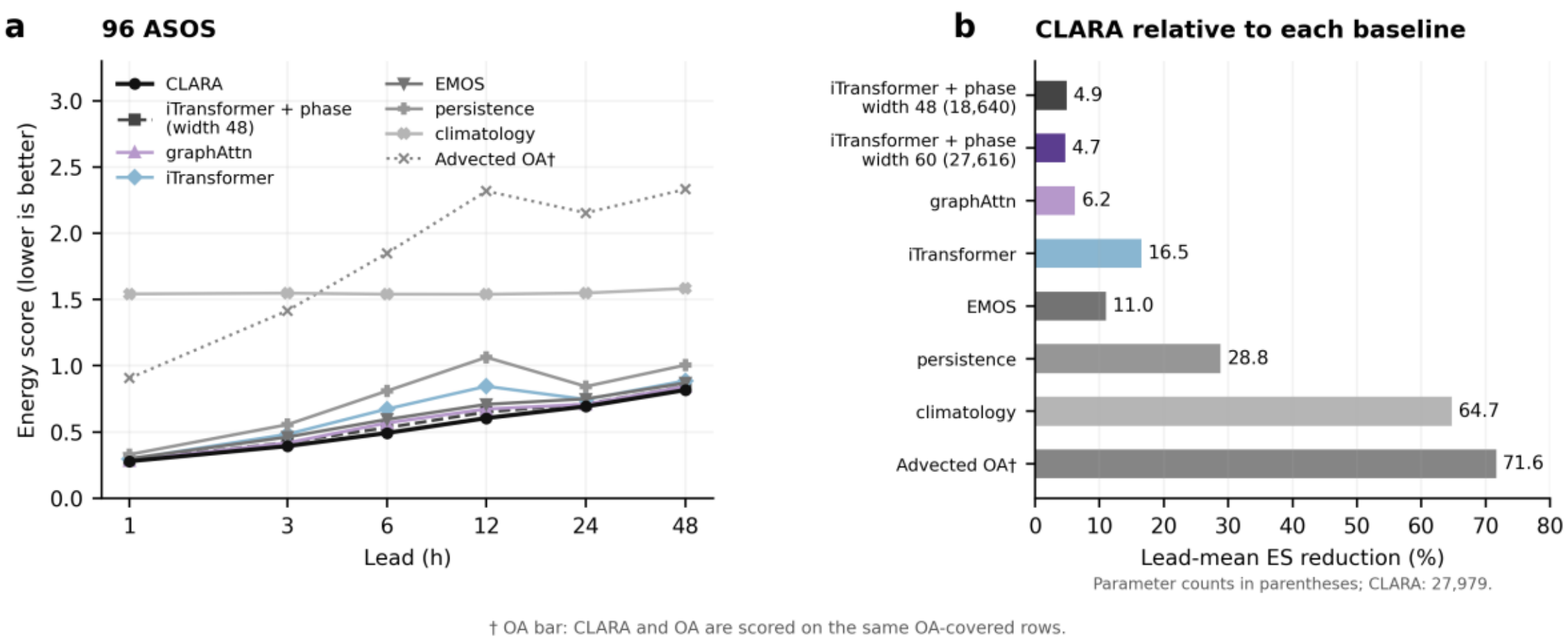

*Figure 2. CLARA improves joint probabilistic forecast skill across all six lead times. Results are for 96 ASOS over six multi-year folds (fold mean of seed means; lower energy score is better).* ***a****, Energy scores of the issued distributions by lead; the advected objective analysis (OA; †) is deterministic and evaluated only on the rows it covers. The CLARA curve uses the full evaluation set.* ***b****, Lead-mean reduction relative to each baseline, 100 ($S_{ref}$ − $S_{CLARA}$)/$S_{ref}$, where each S is the mean energy score across the six leads. For OA, both scores use the same OA-covered rows; the other comparisons use the issued distributions on the full evaluation set. iTransformer + phase receives the same temporal-phase input as CLARA; graphAttn and iTransformer do not receive temporal-phase input. Raw-versus-issued comparisons are in Supplementary Table 4. Values by lead are in Extended Data Table 2 and Supplementary Table 2. In a, iTransformer + phase uses width 48. Panel b additionally includes the width-60 model, retrained to give a parameter count similar to CLARA (27,616 versus 27,979) using the same training, recalibration and evaluation protocol. Parameter counts are shown in parentheses for the two widths. The capacity comparison covers six multi-year folds and eight seeds (Supplementary Note 9).*

Against iTransformer (Liu, Y., et al., 2024) given the same temporal-phase input, the lead-mean improvement is 4.9%, and 4.7% after increasing its width and retraining to give a parameter count similar to CLARA (Fig. 2b; Supplementary Note 9); against the statistical baselines it is 11.0% (EMOS), 28.8% (persistence) and 64.7% (climatology). The advected objective analysis, a deterministic physical baseline scored only where its upwind point lies inside the analysis grid, has a 1 h station-pressure root-mean-square error (RMSE) of 15.2 hPa because of the grid-versus-station elevation gap; on the same rows CLARA's RMSE is lower for all five variables at all six leads (Supplementary Table 1).

Against the adapted observation-based Corrformer (Wu et al., 2023), CLARA has lower energy scores at every lead in all six multi-year folds: 35.3% lower at 1 h and 7.2–9.6% at 12–48 h (Supplementary Note 2, Supplementary Table 3). Among CLARA, iTransformer + phase, EMOS and persistence, CLARA has the lowest fold-mean RMSE in all 30 variable–lead combinations and continuous ranked probability score (CRPS) in 27; exceptions are EMOS for both pressures at 1 h and iTransformer + phase for wind speed at 24 h (Supplementary Fig. 1). Retrained for each 2026 season, CLARA outperforms all five pre-specified baselines at every lead on both networks (60/60 comparisons per network; Supplementary Note 1).

## Joint uncertainty and recalibration

Because the joint covariance is explicit, its calibration can be measured at every lead. Before recalibration the mean squared Mahalanobis distance $\mathbb{E}[d^2]$ rises from 3.1 at 1 h to 5.6 at 24–48 h on the 96-station multi-year folds, crossing the ideal $V = 5$ (the number of variables) between 6 and 12 h (Fig. 3).

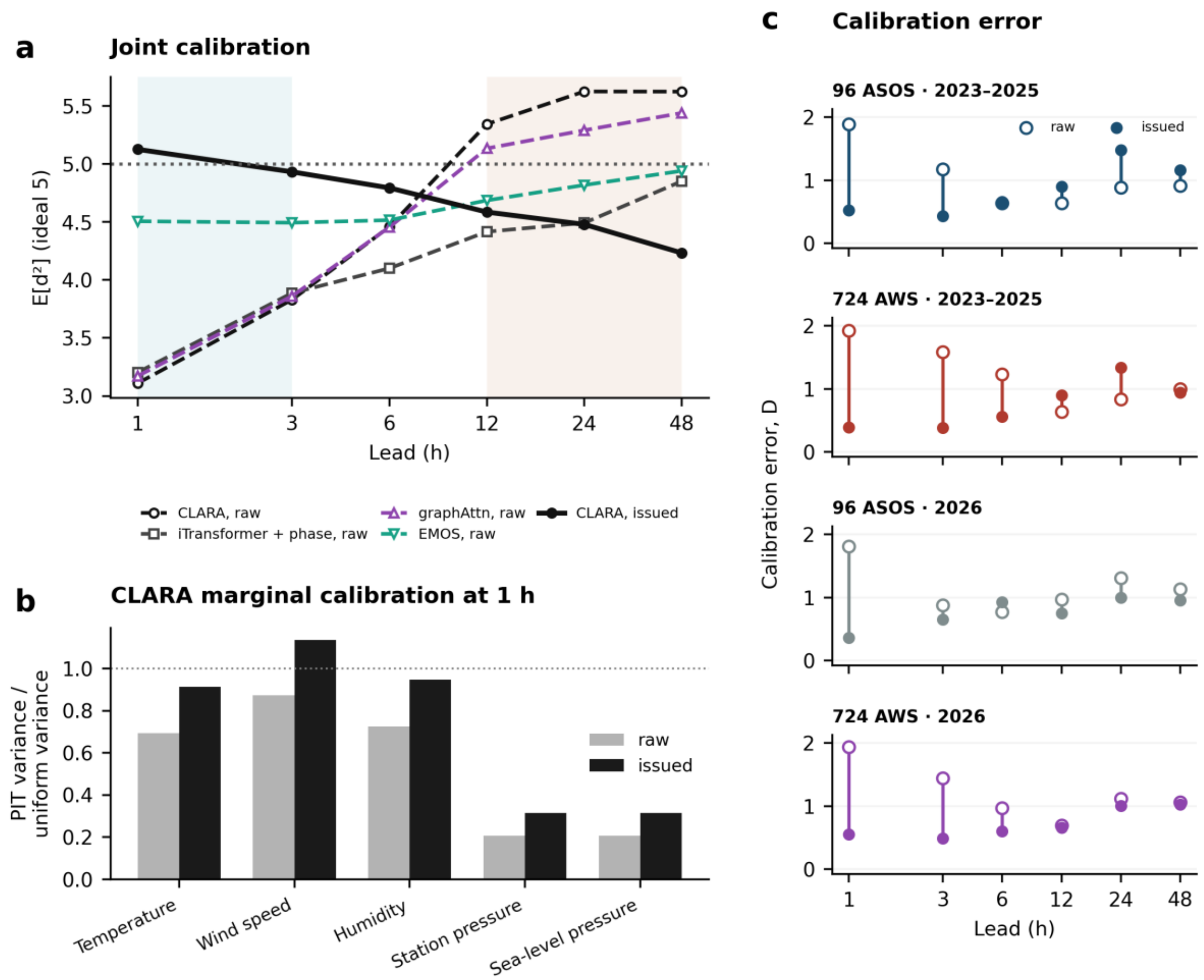

*Figure 3. Explicit joint covariance supports calibration diagnostics before and after lead-wise rescaling.* ***a****, Mean squared Mahalanobis distance E[d²] (ideal 5) for CLARA and three baselines before recalibration, and for issued CLARA forecasts (96 ASOS, multi-year folds). Open markers denote raw distributions; filled markers denote issued CLARA forecasts.* ***b****, For CLARA on 96 ASOS over the multi-year folds, per-variable PIT variance divided by the uniform reference variance at 1 h; the reference value is one.* ***c****, For CLARA, calibration error D = |E[d²] − 5|, computed per run and then averaged over seeds and folds. The upper two rows show 96 ASOS and 724 AWS over three summer and three winter folds in 2023–2025; the lower two rows show the winter and summer 2026 evaluation seasons. Models were trained separately for each evaluation season using preceding data. Open and filled circles show raw and issued forecasts on a common vertical scale. The connecting segment shows the direction and size of the change; rescaling does not improve every network–period–lead combination. The blue and tan backgrounds in a mark 1–3 h and 12–48 h, respectively; they are lead ranges, not confidence intervals or calibration acceptance bands.*

The lead-conditioned scale $s(\tau)$, fitted on a held-out calibration split, reduces the deviation $D = |\mathbb{E}[d^2] - V|$, computed per run and averaged over seeds and folds, at 1–3 h in all six folds, at 1 h from 1.89 to 0.52 and at 3 h from 1.17 to 0.43 (Fig. 3c, Extended Data Table 3), and leaves it nearly unchanged at 6 h (0.63 to 0.66). At 12–48 h, where the uncorrected $\mathbb{E}[d^2]$ already exceeds $V$, it over-corrects: $D$ rises from 0.64–0.91 to 0.90–1.48, with an improvement in only two of the six folds at each lead. The energy score changes by at most 0.0063 (0.8%), and a global scale does not reduce $D$ at any

lead. The 1 h improvement replicates on the 724-station network (1.92 to 0.39), where mean $D$ increases at 12–24 h in the multi-year folds. In 2026, the 1 h reduction is 1.81 to 0.36 on 96 ASOS and 1.94 to 0.55 on 724 AWS; mean $D$ increases only at 6 h on 96 ASOS and at no lead on 724 AWS. Recalibrating all baselines identically leaves the ranking unchanged (Supplementary Table 4). After recalibration with $s(\tau)$, the 1 h probability integral transform (PIT) variance ratios of temperature, wind and humidity are close to one (0.91–1.14), whereas the two pressures remain over-dispersed (0.31–0.32; Fig. 3b), consistent with the diagonal floor of the Cholesky factor (Supplementary Note 4).

Replacing the learned correlations by conditional independence while keeping the marginal variances worsens the joint negative log-likelihood by 1.0–2.8 nats per station and the energy score by 0.1–0.7%, in all six folds at every lead (Fig. 4a,b). A fixed correlation matrix estimated on the training split, however, attains a lower negative log-likelihood than the learned correlations on the 96-station network (by 0.06–0.57 nats) and a lower CRPS for the pressure difference; on the 724-station network the log-likelihood gap is small at 1–24 h and reverses at 48 h (Supplementary Note 1). Lowering the Cholesky diagonal floor from 0.05 to 0.005 gives the learned correlations a lower negative log-likelihood than the fixed ones at 3–48 h but worsens the energy score by 0.8–3.5% relative to the original-floor model (Fig. 4d; Supplementary Note 4).

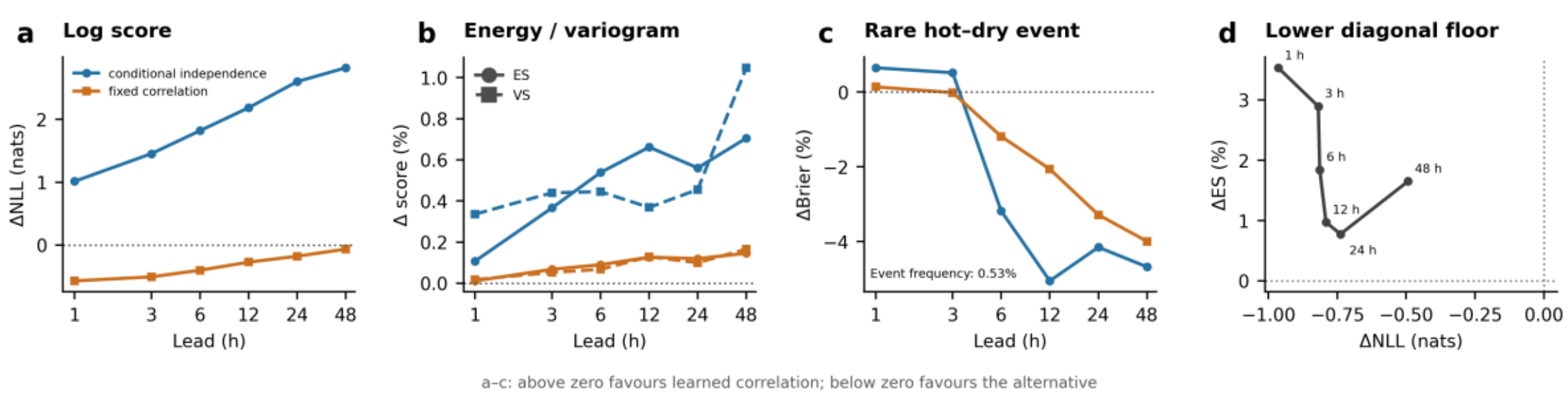

*Figure 4. Predictive dependence and score-specific trade-offs (96 ASOS, six multi-year folds, issued distributions; eight seeds in a–c and three seeds in d). In a–c, differences are alternative minus learned correlation with the marginal distributions held fixed; positive values favour learned correlation. **a**, Changes in joint negative log-likelihood (nats) under conditional independence and under a fixed empirical correlation. **b**, Relative changes in energy score (ES) and variogram score (VS), each expressed as a percentage of its learned-correlation value. **c**, Relative change in Brier score of the rare compound hot–dry event, expressed as a percentage of its learned-correlation value; the observed event frequency is averaged over leads. **d**, Retraining with the Cholesky diagonal floor lowered from 0.05 to 0.005 (seeds 0–2, 2023–2025 folds): change in negative log-likelihood (horizontal axis, lower floor minus original floor, nats) and relative change in energy score (vertical axis, percentage of the original-floor score), by lead.*

## Derived quantities and compound events

The learned correlations change the predictive distribution of quantities that combine variables, by an amount that depends on the quantity (Methods; Extended Data Table 4, Fig. 4). For the pressure difference $\mathrm{ps} - \mathrm{pa}$, removing the correlations increases the mean predictive standard deviation by a factor of 2.7 at 1 h rising to 10.2 at 48 h. Even with the learned correlations, $\mathbb{E}[z^2]$ of the standardized difference $z$ (Supplementary Note 5) is 0.10–0.15 against an ideal of one, so the predictive distribution of this quantity is still too wide; the fixed empirical correlation narrows it to 0.48–0.66 of that spread and yields a lower CRPS for this quantity. Where a single variable dominates, the correlation structure matters little: under conditional independence the PIT variance ratio changes by at most 0.11 for the discomfort index and 0.06 for wind chill. For the rare compound hot–dry event (base rate about 0.5%) the Brier score changes by −5.0% to +0.7%; at 6–48 h both alternatives score better than the learned correlations on this event, by 1.2–5.0% (Fig. 4c).

## Ablation and the routing mechanism

Retraining without one component on the same folds and seeds measures what it adds (Fig. 5c, Extended Data Table 5, Supplementary Fig. 3). Without the temporal-phase input (no-phase) the energy score rises at every lead, most at 6–12 h (by 15.2% and 11.4% on 96 ASOS and by 13.2% and 9.3% on 724 AWS), in at least four of the six folds at every lead and in all six at 3–24 h. Without the advection-routing bias (no-advection) it rises by 0.7–1.3% at 1–12 h and by 0.24–0.25% at 24–48 h on 96 ASOS, in all six folds at 1–12 h; on the denser 724-station network the rise is about twice as large (2.0–2.8% at 1–6 h). Retraining with zero wind in the distance bias produces small, mixed changes in energy score on 96 ASOS (−0.28% to +0.31%) and on 724 AWS (−0.10% to +0.14%), worse in two to five of the six folds on each network. In the strongest-wind decile, changes also differ in sign between folds on both networks (Supplementary Note 1).

Conditioning $\gamma$ on lead adds little beyond a single learned constant: removing the conditioning (constant-γ) raises the energy score by 0.4–0.5% at 1–3 h on 96 ASOS and by 1.3–1.4% on 724 AWS, whereas at 6–48 h the change on 96 ASOS lies between −0.07% and +0.18%, and supplying the lead only to the coupling does not recover what is lost when it is removed from the token embedding (Supplementary Note 1). The learned curve is reproducible across folds and seeds: $\gamma(\tau)$ decays monotonically from 1.61 at 1 h to 0.15 at 48 h on 96 ASOS (47 of 48 runs) and from 3.22 to 0.06 on 724 AWS (all 30 runs). With zero wind it also decays monotonically from 1.60 to 0.16 on 96 ASOS (47 of 48 runs) and from 2.93 to 0.21 on 724 AWS (all 30 runs; Fig. 5a).

## Reading the learned coefficient

Attention weighs the distance term against the content-path scale, which the lead can also change (Lemma 1), so a decaying coefficient alone does not show that the distance term loses weight. On the stored 96-station models the content scale $\lambda$ ranges from 1.21 to 1.88 across leads in the fold mean (a factor of about 1.55; monotone in 15 of 48 runs), while the spread of the distance term relative to it falls from 1.55 at 1 h to 0.12 at 48 h and its effect on the attention weights (Kullback–Leibler divergence) from 0.78 to 0.017 nats, both monotonically in 47 of 48 runs (Fig. 5b; Supplementary Note 3, equation (S1)).

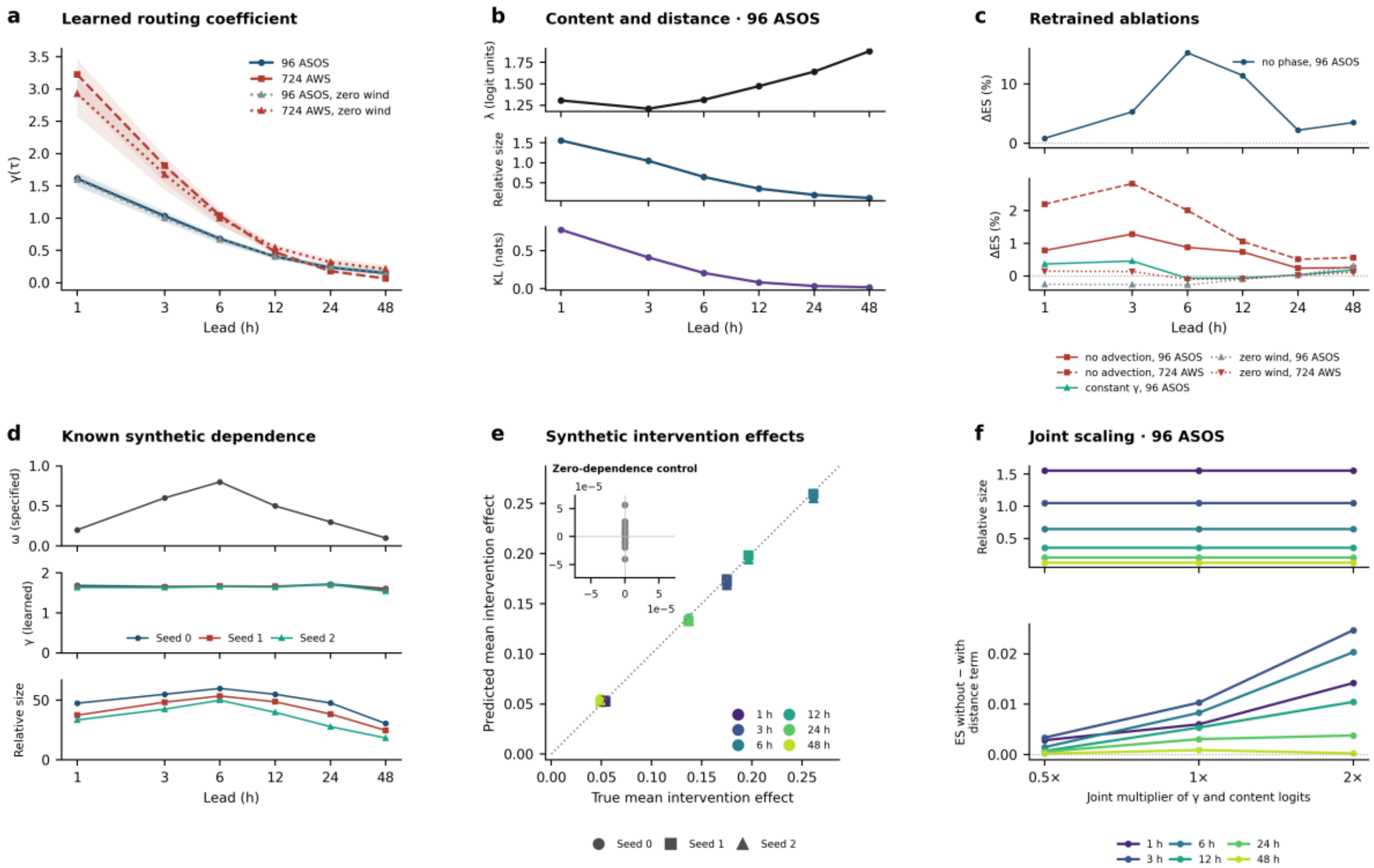

*Figure 5. A learned routing coefficient alone does not measure forecast influence. **a**, Learned γ(τ) for 96 ASOS and 724 AWS, each with and without wind in the distance bias. Lines are means across the six multi-year folds after averaging seeds within each fold; shading spans the minimum and maximum fold means, not a confidence interval. **b**, Content-logit scale λ, relative size of the distance term and Kullback–Leibler divergence between attention weights with and without that term (96 ASOS). The quantities are shown on separate axes in logit units, dimensionless units and nats. Corresponding AWS diagnostics are reported in Supplementary Note 3. **c**, Multi-year energy-score change after retraining ablations, relative to CLARA (%), for the issued distributions; positive values favour CLARA. Temporal phase and routing ablations use separate vertical scales. Results are for 96 ASOS, with the 724 AWS no-advection and zero-wind series also shown. Extended Data Table 5 gives all evaluated network–variant combinations; Supplementary Fig. 3 provides an overview. **d**, Specified generating weight ω, learned γ and relative size in the dependent synthetic task; the specified weight is shared across seeds and shown once in grey, whereas colours and markers distinguish the three seeds for the learned quantities. **e**, Predicted versus true mean intervention effect, with one point per seed and lead; colours identify leads and marker shapes identify seeds. The dotted line is equality. The inset shows the zero-dependence control on an expanded scale. These are means over interventions, not individual intervention effects. **f**, Inference-time intervention on the 48 stored 96 ASOS models (eight seeds across six folds). Jointly multiplying γ and the content logits by 0.5, 1 or 2 leaves the relative size of the distance term unchanged (upper) but changes the energy-score difference, calculated as the score without the distance term minus the score with it (lower). Values are averaged over seeds within each fold and then over folds. Colours identify the six leads. These interventions do not retrain the models; panel c instead compares separately retrained ablations. Supplementary Note 3 reports the full scaling analysis.*

In a synthetic task, each target depends on a kernel-weighted average of upwind inputs with a known lead-dependent weight $\omega(\tau)$ (Methods; Fig. 5d,e). When the five inputs with the largest true kernel weight are perturbed, the effect on the trained model's predictive mean tracks the true effect: across leads and seeds the mean signed error is at most 0.0065 in magnitude and the mean absolute error at most 0.0077, against mean true effects of 0.05–0.26, and the sign is correct in every intervention. The coefficient does not: $\gamma$ stays between 1.54 and 1.72 at every lead while the generating weight $\omega(\tau)$ varies eight-fold, and its ranking of leads agrees only weakly with that of the true effect (rank correlation 0.14–0.54), whereas the ranking based on the relative size of the distance term agrees closely (0.89–1.00). With the dependence removed ($\omega \equiv 0$) the model's mean effect is below 0.0001 in magnitude and its mean absolute effect below 0.00044 at every lead, yet in one seed $\gamma$ reaches 24 at 48 h and the Kullback–Leibler divergence of the attention weights 16 nats: a large coefficient and a large change in attention occur with a negligible effect on the forecast.

Relative size tracks the ordering of effects in this task but does not determine them. On the 48 stored 96-station models, jointly halving, retaining or doubling $\gamma$ and the content logits at inference preserves their ratio. The energy score without the distance term minus that with it is 0.0015, 0.0083 and 0.0203, respectively, at 6 h (Fig. 5f; Supplementary Note 3).

## Replication across observing networks

On the denser 724-station Automatic Weather Station (AWS) network, CLARA has lower fold-mean energy scores than all tested baselines at every lead and lower scores in all 108 comparisons with the three statistical baselines. Against iTransformer with the same temporal-phase input, the lead-mean energy score is 7.0% lower (Supplementary Table 5, Extended Data Fig. 1), and 6.2% lower with the wider, similarly sized comparator (Extended Data Fig. 1b; Supplementary Note 9). On the rows covered by the advected OA, it is 65.4% lower than that of OA; coverage falls from 79% at 1 h to 31% at 48 h (Supplementary Fig. 2).

On stored 724-station models, half the input stations remained fixed as evaluation stations while 0–100% of the remaining context stations were removed, without retraining (three seeds, six folds; Extended Data Fig. 2a). Removing every context station raises the energy score at the evaluation stations by 0.3–0.7%. Two case forecasts chosen by rules fixed before inspection illustrate the 6 h temperature uncertainty (Extended Data Fig. 2b,c; means over five seeds). In the summer case the spread of the error (1.82 °C) is close to the predicted standard deviation $\sigma$ (mean 1.90 °C) and the central 90% interval covers 92% of the stations. In the winter cold-surge case a domain-wide warm bias (+1.89 °C) dominates the error, the predicted $\sigma$ (mean 1.61 °C) is smaller than the error spread (1.97 °C), and the interval covers 60% of the stations.

## Replication across regional observing networks

Models with the same configuration were trained separately in ten regions on six continents (Köppen A–D), using up to 50 stations per region from the NOAA Global Historical Climatology Network hourly (GHCNh) archive. Regional models predict three-variable joint distributions because pressure reporting varies regionally (Methods). CLARA outperforms persistence in all 60 region–lead combinations, in each of the six multi-year folds (Fig. 6c, Extended Data Table 6), and it improves on climatology in all 60 combinations, by 47.6–78.0% at 1 h and by 30.2–47.6% at 48 h. Against iTransformer + phase with a similar parameter count, the multi-year lead-mean energy score is 2.8–5.4% lower in each region; CLARA has a lower score in 57 of 60 region–lead means (Fig. 6d; Supplementary Note 12). In both 2026 seasons, CLARA also outperforms persistence in every region–lead combination (Supplementary Note 12).

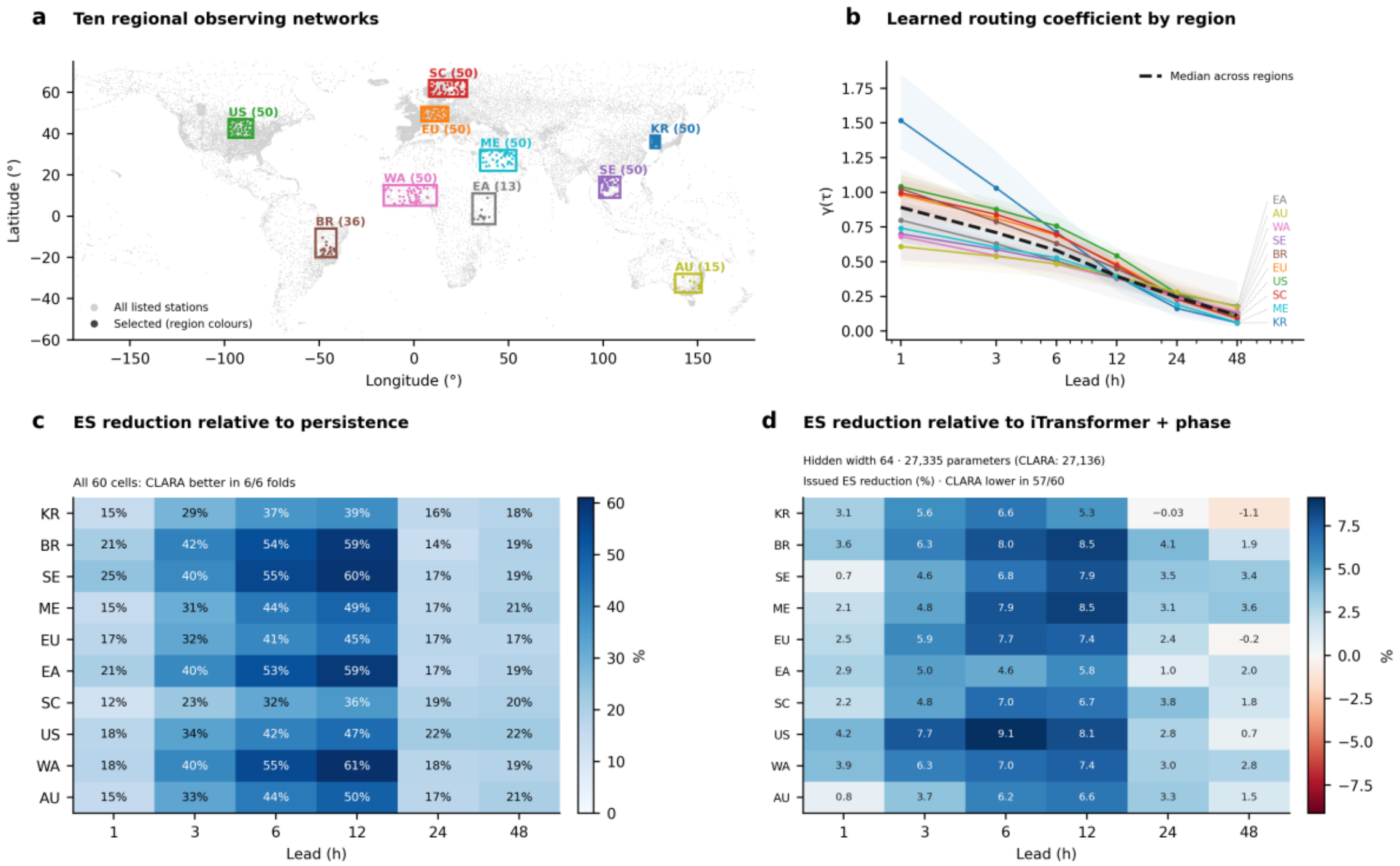

*Figure 6. Forecast skill and lead-dependent routing replicate across ten regional observing networks. Results use six multi-year folds read in UTC and three seeds per region; the predicted variables are temperature, wind speed and humidity.* ***a****, Selected stations (region colours) over the GHCNh station list (grey); parentheses give the number selected. The same region colours are used in b.* ***b****, Learned routing coefficient γ(τ): regional means and minimum–maximum ranges across the 18 runs. Region codes at the line ends are offset vertically for legibility and connected to their corresponding values. The dashed black line is the median of the ten regional means at each lead.* ***c, d****, Issued energy-score reduction relative to persistence (c) and iTransformer + phase (d), calculated as 100 ($S_{reference}$ − $S_{CLARA}$)/$S_{reference}$ from scores averaged over seeds within each fold and then over folds. Regions are ordered by increasing median nearest-neighbour distance among the selected stations. Colour scales differ between c and d; positive values favour CLARA. Against persistence, CLARA has a lower score in all six folds for every region–lead combination. The regional iTransformer + phase uses hidden width 64 and has 27,335 trainable parameters, compared with 27,136 for CLARA, and follows the same training, recalibration and evaluation protocol (Supplementary Note 12). CLARA has a lower score in 57 of the 60 region–lead means against this comparator. Region codes: KR, Korean Peninsula; EU, central Europe; US, US Midwest; SC, Scandinavia; SE, Southeast Asia; BR, central Brazil; WA, West Africa; EA, East Africa; AU, southeastern Australia; ME, Middle East.*

The learned routing coefficient $\gamma(\tau)$ decays monotonically in the mean for each of the ten regions and in 179 of the 180 runs (Fig. 6b); its level, however, is region-specific. $\gamma(1\,\mathrm{h})$ ranges from 0.61 over southeastern Australia to 1.52 over the Korean Peninsula and follows station spacing: across the twelve networks of this study, the median nearest-neighbour distance and the nominal decay length $100\,\mathrm{km}/\gamma(1\,\mathrm{h})$ of the distance factor correlate at 0.81.

Two operational NWP references were compared on the stations and times they share with CLARA, as a diagnostic rather than a like-for-like benchmark: neither is post-processed to stations, whereas CLARA is trained and recalibrated on them. Over the US Midwest (17 shared stations, two folds) CLARA has a lower RMSE than the High-Resolution Rapid Refresh (HRRR) for all three variables at 1 h and for humidity at every lead, while HRRR is more accurate for temperature and wind from 3 h (Supplementary Table 6). Over the Korean Peninsula (30 shared stations, six folds), CLARA achieves a 22–34% lower CRPS than the raw European Centre for Medium-Range Weather Forecasts (ECMWF) ensemble for all three variables at 6–12 h, in all six folds, whereas the ensemble is better for temperature from 24 h (Supplementary Table 7, Extended Data Fig. 3).

## Discussion

Learning a full predictive covariance makes the value of cross-variable dependence measurable, and the measurement depends on the scoring rule. Against conditional independence the learned correlations improve the joint log-likelihood by one to three nats per station but the energy score by less than 1%, consistent with the limited sensitivity of the energy score to dependence (Scheuerer & Hamill, 2015) and the sensitivity of the log score to over-narrow predictive covariances. A fixed empirical correlation nevertheless achieves a better log-likelihood and pressure-difference CRPS on the 96-station network. The floor diagnostic implicates the Cholesky diagonal floor in the excessive predictive spread of the nearly collinear pressures: lowering it reverses the log-likelihood ranking at 3–48 h and reduces the 1 h pressure CRPS below that of EMOS, at the cost of a worse energy score. The floor therefore sets a trade-off between the joint likelihood and the kernel scores, and the lead-wise recalibration, which by Proposition 1 cannot alter the correlations, corrects the scale at short leads but over-corrects at long ones. Conformal calibration can complement this approach by calibrating joint prediction regions, but coverage alone does not determine the dependence structure or the distributions of derived quantities.

On both networks, the small effects of zero-wind retraining, contrasted with the losses without the distance term, indicate that routing benefits come mainly from distance weighting. The no-advection loss is about twice as large on the 724-station network, where network type and density change together, and the learned decay length of the distance factor follows station spacing across twelve networks. The median surface wind moves the upwind location by 5–6 km in an hour, a small fraction of the 31–62 km decay length of the distance factor at 1 h on the two nationwide networks, and at long leads, where the displacement is larger, the coefficient is small. The bias thus selects among nearby stations; surface-wind displacement adds no consistent benefit in either zero-wind comparison.

The decrease of $\gamma(\tau)$ with lead is reproduced across years, networks, seeds and ten regions, including models trained with zero wind in the distance bias on both domestic networks; it corresponds to a longer nominal decay length of the distance factor, while the effective attention footprint also depends on the content logits and station geometry. Lemma 1 applies to any attention with a learned additive-bias coefficient whose conditioning variable also enters the content path. Such a coefficient can be read only together with the content scale, and a claim about influence requires an intervention on the forecast: in the synthetic task $\gamma$ was nearly constant while the generating weight varied eight-fold. Joint scaling on the weather models further shows that relative weight alone does not determine forecast influence (Fig. 5f).

The study has four limitations. (i) The evaluation outside Korea uses three variables, because pressure reporting varies by region; what is reproduced across regions is the declining form of $\gamma(\tau)$ and the multi-year lead-mean advantage over persistence, climatology and the similarly sized learned comparator, not the level of $\gamma$ (Fig. 6b–d). (ii) The NWP references cover two regions, shared stations only and products as disseminated; CLARA does not inherit NWP latency, and combination with post-processing is future work. (iii) Parameter-count-controlled learned comparisons cover one model family, iTransformer; some regional long-lead means favour this comparator. (iv) The joint distribution is Gaussian: the assumption yields the closed-form likelihood and analytic, correlation-preserving recalibration, but a Gaussian cannot represent multimodal, asymmetric or heavy-tailed conditionals; violations are diagnosed by the PIT (Fig. 3).

CLARA jointly learns five-variable means and full covariances from station observations for 1–48 h forecasts, with CPU training and prediction and no NWP input. The explicit covariance makes calibration measurable at every lead and correctable analytically under stated conditions, and it shows that the value of learned dependence differs between scoring rules. The learned routing coefficient decays with lead across networks and regions; its interpretation requires the content scale and measured forecast interventions. Multi-year evaluations show lower energy scores than statistical baselines on both nationwide networks and persistence and climatology in ten regions, and lower lead-mean scores than similarly sized learned models with matched temporal inputs in both networks and every region. One forward pass per lead provides marginal quantiles and exceedance probabilities in closed form, supporting low-latency, NWP-independent prediction.

## Methods

### Related approaches

Lead conditioning in a multi-lead network is usually supplied as an extra input encoding of the lead (Sønderby et al., 2020; Espeholt et al., 2022); where in the network the lead enters affects performance (Nguyen et al., 2024), and attention has been used to interpret what a spatio-temporal model relies on (Liang et al., 2024). Recent systems share one model across leads (Bashir et al., 2026; Liu & Rahnemoonfar, 2026), but their forecasts are deterministic. Observation-centred systems combine station data with radar and NWP (Miralles et al., 2025, 2026), extend range from radar and satellite (Sønderby et al., 2020; Espeholt et al., 2022; Andrychowicz et al., 2023) or reduce the dependence on reanalysis (Allen et al., 2025; Pinnington et al., 2026). A probabilistic extension of such a system has also been proposed (Almeida et al., 2026). These systems nonetheless use assimilated fields or issue marginal or deterministic forecasts, and a probabilistic loss is left as future work (Pinnington et al., 2026).

Spatial coupling in attention has been informed by distance, position or geostatistical covariance (Calleo, 2025), while graph attention learns neighbour importance from the data alone (Veličković et al., 2018; Lam et al., 2023). Displacement itself has been handled outside the attention — by learned advection operators (de Bézenac et al., 2019; Verma et al., 2024), by location-variant recurrent structure (Shi et al., 2017) or by optical-flow nowcasting (Ayzel et al., 2019). iTransformer attends over the variable axis (Liu, Y., et al., 2024). Adding a learnable bias to an attention logit while the same conditioning variable also enters the content path makes the effect of the bias coefficient depend on the content scale, so the coefficient alone does not fix the relative weight of the two paths. Work on attention identifiability has posed such questions for the relation between weights and outputs rather than for a conditioned coefficient (Brunner et al., 2020), and the length-extrapolation literature treats such a bias as a scaling device, not a coefficient to interpret (Press et al., 2022). Post hoc attribution in time series raises the same question of when an interpretable quantity can be trusted: methods disagree on the same task, and metrics to adjudicate them have been proposed (Turbé et al., 2023).

Predictive distributions are built by generating members (Lang et al., 2024; Pacchiardi et al., 2024; Lakatos, 2026; Alet et al., 2025), by post-processing (Gneiting et al., 2005; Rasp & Lerch, 2018; Schefzik et al., 2013; Baran & Möller, 2017; Asch et al., 2026; Wessel et al., 2024; Assouline et al., 2026; Höhlein et al., 2024) or by emitting the parameters directly (Salinas et al., 2019; Wen & Torkkola, 2019; Verma et al., 2024; Muschinski et al., 2024); standard benchmarks presuppose a supplied ensemble (Demaeyer et al., 2023). CLARA follows the third route and extends it so that the parameters reflect neighbouring stations and the lead. In CLARA the emitted Gaussian parameters give the joint predictive law in closed form, so one forward pass per lead yields the joint forecast. Extended Data Table 1 compares prior systems by input data, output and resolution, lead range, probabilistic output, lead-time handling and reliance on NWP or reanalysis; among the systems compared, and in the wider survey of Supplementary Note 6, we find none that issues a joint predictive distribution from station observations alone while modulating both the strength of an upwind-position bias and the magnitude of the covariance by lead.

## Forecast problem

The prediction target is the joint probability distribution of several surface variables at each observing station. Let $t$ be the base time, $\tau$ the lead time and $\mathcal{S}_t$ the set of stations available at that time. At each target station $i \in \mathcal{S}_t$ the predictand at target time $t + \tau$ is the five-dimensional vector

$$y_{i,t+\tau} = \left(\mathrm{ta}_{i,t+\tau},\ \mathrm{ws}_{i,t+\tau},\ \mathrm{hm}_{i,t+\tau},\ \mathrm{pa}_{i,t+\tau},\ \mathrm{ps}_{i,t+\tau}\right)^{\top} \quad (1)$$

where $\mathrm{ta}$, $\mathrm{ws}$, $\mathrm{hm}$, $\mathrm{pa}$ and $\mathrm{ps}$ denote temperature, wind speed, relative humidity, station pressure and sea-level pressure. CLARA issues the conditional joint predictive distribution of this vector for each target station and lead. The distribution represents dependence among variables at one station; it is not a spatial joint distribution linking the future states of different stations.

The inputs are, for each station $j \in \mathcal{S}_t$ providing data, the current observation $y_{j,t}$, the tendency over the preceding 60 min $\Delta y_{j,t}$, static features $g_j$ describing the station (elevation in the domestic experiments; elevation, latitude and longitude in the regional experiments), the lead time $\tau$, and the temporal phase of base and target time $\phi_{t,\tau}$.

One model is trained jointly over the six leads of 1, 3, 6, 12, 24 and 48 h. The forecast at each lead is computed directly from the observations at the same base time; the forecast at one lead is never used as input at the next.

Station pressure and sea-level pressure are strongly correlated by construction. Both are retained as predictands to test how well the model represents the joint uncertainty of strongly dependent variables.

CLARA uses no NWP, reanalysis or gridded field at any stage of training or inference. The dynamic information entering a forecast is restricted to station observations; only terrain and station characteristics enter as static information.

## Model overview

CLARA is an attention-based model that issues a joint predictive distribution over several variables at each station from observing-network information alone. It has four stages: representation of the inputs, spatial combination across stations, construction of the joint predictive distribution, and lead-wise recalibration of its uncertainty (Fig. 1).

Each station's inputs are first mapped to a single station token containing the current observation, the 60-min tendency, static station features, the lead time and the temporal phase. Inputs with different units and ranges are preprocessed, standardized and projected to a latent vector of fixed dimension, which serves as the representation of that station in the attention that follows.

The temporal phase supplies the diurnal cycle and the seasonal variation. The diurnal phases of base and target time are each encoded as a sine–cosine pair, and the day-of-year seasonal phase likewise, giving six cyclic features: two for the base-time diurnal phase, two for the target-time diurnal phase and two for season. The base-time phase indicates the position of the base time within the diurnal cycle, the target-time phase identifies the hour being predicted, and the seasonal phase identifies the time of year.

In the default configuration the six temporal features are concatenated with the observations, tendencies, static attributes and lead, and a single embedding layer maps them to the station token, so that time is processed in the same representation space as the observations.

Station tokens are then combined with information from neighbouring stations by multi-head self-attention (Vaswani et al., 2017), with a lead-conditioned bias added to the usual attention score to control how strongly neighbours near the upwind position are weighted. In CLARA, advection routing denotes this lead-conditioned attention bias based on distance to a wind-derived upwind location: the location is obtained by moving the station against the base-time wind on the sphere, not by tracing an air parcel; below it is called the advection bias. The bias and its lead-wise coupling strength are defined below.

The spatial information returned by the attention is merged with the existing station token in a proportion set by a sigmoid residual gate. This residual connection preserves the original input while adding the spatial information the forecast requires, and a feed-forward network then transforms the merged information into the station representation used downstream.

Finally, the joint output layer maps the updated representation to a mean vector and a Cholesky factor for the five-variable covariance at each station. After training, a lead-wise scale estimated on a separate calibration split is applied to the covariance to recalibrate the magnitude of the predictive uncertainty.

## Lead-conditioned advection-routing attention

To let the attention follow the transport of weather by the wind, CLARA computes a lead-dependent upwind position. With $p_i$ the position of target station $i$ and $w_i$ the wind vector observed at base time, the upwind position for lead $\tau$ is the point reached from $p_i$ by moving a distance $\|w_i\|\tau$ against the direction of $w_i$ along the great circle on a sphere of radius 6,371 km (the spherical destination formula), and station separations are great-circle distances, so the same formula applies at every latitude.

Let $d_{ij}^{\mathrm{up}}$ be the great-circle distance in km between that upwind position and neighbouring station $j$. CLARA feeds this distance into the attention so that stations near the upwind position receive relatively more weight.

Ordinary self-attention scores each neighbour by comparing observational features; write that content score $\ell_{ij}^{\mathrm{content}}(\tau)$. CLARA subtracts a term proportional to the upwind distance,

$$\ell_{ij}^{\mathrm{CLARA}}(\tau) = \ell_{ij}^{\mathrm{content}}(\tau) - \gamma(\tau)\,\frac{d_{ij}^{\mathrm{up}}}{100} \tag{2}$$

where the constant 100 of equation (2) sets the numerical size of the distance term, and $\gamma(\tau)$ is the lead-wise coupling strength defined in equation (3). A salience offset $b_j$, computed per neighbour and head from the neighbour's token, is also added to the logit (Supplementary Note 8) and enters equation (4) through the content logit $e_{ij}$. The expression therefore does not advect the state from upwind to downwind, but reweights the attention so that observations near the upwind position receive more weight. The distance penalty is smaller for neighbours closer to the upwind position.

The strength with which the upwind distance is used is not fixed across leads but is a learnable function of lead,

$$\gamma(\tau) = \mathrm{softplus}(\mathrm{MLP}(\tau_{\mathrm{norm}})) \tag{3}$$

in which a multilayer perceptron (MLP) takes the lead as input and learns that strength. The lead enters as $\tau_{\mathrm{norm}} = (\ln\tau - \ln 360)/0.9$ with $\tau$ in minutes, a log transform centred and scaled by fixed constants so that a wide range of values is handled stably, and the softplus keeps $\gamma(\tau)$ positive.

With the content logits and upwind distances fixed, increasing $\gamma(\tau)$ increases the relative attention odds of nearer neighbours over farther ones.

The upwind position and the attention scores at every lead come directly from the base-time observations.

The lead $\tau$ enters the station token as well as $\gamma(\tau)$, so the change of attention with lead cannot be attributed to $\gamma(\tau)$ alone. The relative contribution of the distance term depends on its size compared with the spread of the content logits over neighbours. To make that spread explicit, write the content logit of query $i$ in one head, including the salience offset, as $e_{ij} = \ell_{ij}^{\mathrm{content}}(\tau) + b_j$ with $\ell_{ij}^{\mathrm{content}}(\tau) = q_i^{\top} k_j / \sqrt{d_k}$, where $q_i, k_j$ are the query and key vectors of dimension $d_k$; let $\lambda_i = \mathrm{SD}_j(e_{ij})$ be its standard deviation over neighbours $j$, the content scale, and $\tilde{C}_{ij} = (e_{ij} - \bar{e}_i)/\lambda_i$ the standardized content score, with $\bar{e}_i$ the mean over $j$. Because a softmax is unchanged when the same constant is added to every logit of a row,

$$\mathrm{softmax}_j\big(e_{ij} - \gamma A_{ij}\big) = \mathrm{softmax}_j\big(\lambda_i\big[\,\tilde{C}_{ij} - (\gamma/\lambda_i)A_{ij}\big]\big) \tag{4}$$

where $A_{ij} = d_{ij}^{\mathrm{up}}/100$ is the scaled upwind distance. The content scale $\lambda_i$ acts as an inverse temperature, and the weight of the upwind distance relative to the content path is $\gamma/\lambda_i$.

In discrete-choice theory the scale parameter of a multinomial logit is not identified and must be normalized (Train, 2009). Softmax attention has the form of a multinomial logit over keys, but its logits carry no unobserved error term, so $\lambda_i$ need not be normalized away: it is computed directly from the logits, and the two factors can be measured separately. We therefore report, on the stored models, the content scale $\lambda$, the size of the distance term relative to it, $\mathrm{SD}_j(\gamma A_{ij})/\lambda_i$, and the Kullback–Leibler divergence of the attention weights from those with the distance term removed, each averaged over queries, heads and samples at every lead (Supplementary Note 3).

**Lemma 1 (scale dependence of the advection coefficient).** For a query row $i$ whose content logits $e_{ij}$ have standard deviation $\lambda_i > 0$ over neighbours $j$, the attention weights are $\mathrm{softmax}_j\big(\lambda_i[\tilde{C}_{ij} - (\gamma/\lambda_i)A_{ij}]\big)$ with $\tilde{C}_{ij}$ of unit spread, so the weight of the upwind distance relative to the content path is $\nu_i = \gamma/\lambda_i$. Hence, for any two leads, $\gamma(\tau_2)/\gamma(\tau_1) = [\nu_i(\tau_2)/\nu_i(\tau_1)][\lambda_i(\tau_2)/\lambda_i(\tau_1)]$: a change of $\gamma(\tau)$ with lead reflects a change of the relative weight only insofar as the content scale does not change with lead.

## Proof of Lemma 1: scale dependence of the advection coefficient

A softmax is unchanged when the same constant is added to every logit of a row, so replacing $e_{ij}$ by $e_{ij} - \bar{e}_i$ leaves the weights unchanged. Writing $e_{ij} - \bar{e}_i = \lambda_i \tilde{C}_{ij}$ and factoring $\lambda_i > 0$ out of the row gives equation (4), with the constant $1/100$ absorbed into $A_{ij}$.

Inside the bracket $\tilde{C}_{ij}$ has unit standard deviation over $j$ by construction, so the distance term enters with coefficient $\nu_i = \gamma/\lambda_i$ against a content term of fixed spread, and $\lambda_i$ multiplies the whole bracket as an inverse temperature. The identity $\gamma = \lambda_i \nu_i$, evaluated at two leads and divided, gives the stated ratio. Both factors are functions of the logits and are therefore computable: $\lambda_i$ as the spread of $e_{ij}$ over $j$, and $\nu_i$ as the coefficient of $A_{ij}$ in the standardized logit.

## Multivariate joint predictive distribution and training loss

The station tokens updated by the attention block enter a joint output layer that produces a mean vector and a covariance matrix for each station and lead. The mean vector carries the predictive centre of temperature, wind speed, humidity, station pressure and sea-level pressure. The covariance matrix specifies the marginal variances and pairwise covariances of the five variables.

The mean at target station $i$ is formed by adding a predicted increment to the base-time observation,

$$\mu_i(t,\tau) = y_{i,t} + \delta_i(t,\tau) \tag{5}$$

where $y_{i,t}$ is the five-vector observed at base time $t$ and $\delta_i(t,\tau)$ is the expected change from $t$ to $t + \tau$. The model therefore learns a departure from persistence rather than predicting the target state directly. The increment $\delta_i(t,\tau)$ is produced by the model and is distinct from the 60-min tendency supplied as input.

The covariance is produced through a Cholesky factorization,

$$\Sigma_i(t,\tau) = L_i(t,\tau)\, L_i(t,\tau)^\top \tag{6}$$

Here $L_i(t,\tau)$ is the Cholesky factor of the covariance. For a five-variable joint distribution the output layer produces the fifteen elements of a lower-triangular matrix; constraining its diagonal to be positive makes the $\Sigma_i(t,\tau)$ of equation (6) a valid covariance matrix. The output transformation of the Cholesky factor and the settings used for numerical stability are given under 'Implementation and training settings'.

CLARA uses the five-dimensional Gaussian with parameters $\mu_i(t,\tau)$ and $\Sigma_i(t,\tau)$ as the joint predictive distribution at each station, from which both the marginal distribution of each variable and the correlation structure follow.

Training minimizes the multivariate Gaussian negative log-likelihood (NLL) of the observed five-vector,

$$\mathcal{L}_i(t,\tau) = 1/2\left(y_{i,t+\tau} - \mu_i(t,\tau)\right)^\top \Sigma_i(t,\tau)^{-1}\left(y_{i,t+\tau} - \mu_i(t,\tau)\right) + 1/2\,\mathrm{logdet}\Sigma_i(t,\tau) + V/2\,\mathrm{log}2\pi \tag{7}$$

averaged over stations, base times and leads, where $V = 5$ is the number of variables ($V = 3$ in the three-variable GHCNh experiments). The first term standardizes the error of the mean by the predicted covariance and the second penalizes over-dispersion, so the error of the mean and the scale and correlation of the uncertainty are learned in one loss. The loss is a strictly proper log score, so minimizing it drives the predictive distribution towards the Gaussian $q$ that minimizes $\mathrm{KL}(p \parallel q)$, with $p$ the conditional distribution of the data. After training, the covariance of equation (6) is used to diagnose calibration by lead and to recalibrate the magnitude of the uncertainty as described next.

### Lead-wise recalibration of the covariance scale

The width of the predictive distribution may be inappropriate at some leads. To correct this, a lead-wise covariance scale is estimated on a separate calibration split once training is complete. The procedure changes only the overall magnitude of the covariance, leaving the predictive mean and the correlation structure untouched. The estimator itself is the multivariate, lead-wise form of the moment matching used in σ-scaling of regression uncertainty (Levi et al., 2022; Laves et al., 2020) and in covariance-inflation estimation for ensemble data assimilation (Li et al., 2009). The same held-out calibration design is used for CLARA's domestic and regional forecast evaluations, including the regional runs that enter the operational NWP reference comparison (Supplementary Note 9).

The recalibrated covariance and the scale at lead $\tau$ are

$$\Sigma_i' = s(\tau)^2\, \Sigma_i, \qquad s(\tau) = \sqrt{\widehat{\mathbb{E}}_{\mathrm{calib}}[d^2]/V} \tag{8}$$

where $\Sigma_i$ and $\Sigma_i'$ are the covariance before and after recalibration and $V$ is the number of variables in the joint distribution. The squared Mahalanobis distance $d^2 = (y - \mu)^\top \Sigma^{-1}(y - \mu)$ standardizes the difference between observation and predictive mean by the predictive covariance, and $\widehat{\mathbb{E}}_{\mathrm{calib}}[d^2]$ is its sample mean on the calibration split. The scale $s(\tau)$ is estimated per lead on that set and applied to an evaluation set held apart from it; the last 15% of each training window serves as the calibration split and is excluded from training.

By equation (8) each variance is multiplied by $s(\tau)^2$ and each standard deviation by $s(\tau)$. Because the same positive scale multiplies every entry of the covariance, the correlation matrix is unchanged. The procedure therefore adjusts only the magnitude of the predictive uncertainty per lead while preserving the dependence structure the joint distribution has learned. The forecast issued at station $i$ and lead $\tau$ is thus the closed-form predictive law

$$y_{i,t+\tau} \ \sim \ \mathcal{N}(\mu_i(t,\tau),\ s(\tau)^2\,\Sigma_i(t,\tau)) \tag{9}$$

in which the mean and the covariance come from equations (5), (6) and (8): a full joint distribution over the five variables, available in one forward pass per lead.

**Proposition 1 (correction of a scalar covariance misspecification).** Suppose the true conditional distribution is Gaussian with the predicted mean $\mu(x)$ and covariance $c(\tau)^2\Sigma(x)$, where $c(\tau) > 0$, $x$ denotes the conditioning inputs and $\Sigma(x)$ the predicted covariance. Then the expected squared Mahalanobis distance is $\mathbb{E}[d^2] = c(\tau)^2 V$, and if the calibration statistics are weakly dependent across sampled days (assumption (A4) of Supplementary Note 11), $s(\tau)^2$ in equation (8) converges in probability to $c(\tau)^2$ as the number of sampled days grows. If $s(\tau) = c(\tau)$, the recalibrated expected squared Mahalanobis distance equals $V$; and for any positive $s(\tau)$ the original correlation matrix is preserved.

If the model is correctly specified, $d^2$ follows a chi-squared distribution with $V$ degrees of freedom, so applying that distribution function to $d^2$ gives a value uniform on the unit interval — the property used for the multivariate probability integral transform diagnostic (Supplementary Note 13). Fig. 3 shows the mean squared Mahalanobis distance and per-variable PIT variance ratios. However, $\mathbb{E}[d^2] = V$ is necessary but not sufficient for the joint distribution to be well calibrated, so the per-variable transform is examined alongside the multivariate one.

**Corollary 1 (per-variable diagonal extension).** Under assumptions (A1), (A2) and (A4) of Supplementary Note 11, if the true covariance is $R^{1/2}\Sigma R^{1/2}$ with $R = \operatorname{diag}\kappa_v(\tau)$ and $\kappa_v(\tau) > 0$ — correct correlations but a separate scale error for each variable — then the per-variable scales $s_v(\tau)^2 = \widehat{\mathbb{E}}_{\text{calib}}[z_v^2]$, with $z_v = (y_v - \mu_v)/\sqrt{\Sigma_{vv}}$, converge in probability to $\kappa_v(\tau)$, and at the population value the recalibrated covariance $\Sigma^S = S\Sigma S$ with $S = \operatorname{diag} s_v(\tau)$ equals the true covariance; the proof is given in Supplementary Note 11.

**Proof outline.** Under assumptions (A1)–(A2) of Supplementary Note 11, the squared Mahalanobis distance divided by $c(\tau)^2$ has a chi-squared distribution with $V$ degrees of freedom. Under the bounded day-block size and mixing condition in (A4), the variance of the scale-squared estimator is $O(1/B)$, where $B$ is the number of sampled day blocks, giving consistency by Chebyshev's inequality. Assumption (A3) transfers the calibration-period scale to evaluation. At the population value $s(\tau) = c(\tau)$, recalibration recovers the stated Gaussian covariance, while any positive scalar rescaling preserves correlations. The full proof and the treatment of general misspecification are given in Supplementary Note 11.

**Remark.** Where the predictive mean is biased or the misspecification is not a scalar, the same statistic tracks a sample-averaged pseudo-true scale that absorbs the bias, and converges to a fixed limit if that average converges; Supplementary Note 11 derives that pseudo-true scale under general misspecification (equation (S5)) and provides the multivariate probability-integral-transform result, a trace-based counterexample showing that $\mathbb{E}[d^2] = V$ is necessary but not sufficient, the proof of Corollary 1 and the numerical verification of these identities.

## Observations and preprocessing

Two domestic station networks and one international archive serve different purposes (Extended Data Table 7). The 96 stations of the Automated Synoptic Observing System (ASOS) are the primary evaluation set. Their record from 2021 is split into eight seasonal rolling-origin folds: six multi-year folds, summer (JJA) and winter (DJF) of 2023, 2024 and 2025, and two held-out folds, December 2025–February 2026 and June–August 2026. The configuration of CLARA was chosen in earlier experiments on 2024–2025 data from AWS stations in central Korea, which overlap the 2024 and 2025 folds in time; the 2026 folds were not used to select CLARA's configuration or the pre-specified primary comparisons. Each fold trains on the 23–29 months before its evaluation season, holding out the last 15% of that window for calibration, and evaluates base times every four days at 00, 06, 12 and 18 Korea Standard Time (KST): 92 scheduled base times per season and 548–552 retained evaluation groups on 96 ASOS. A group is a base-time–lead pair with both times inside the evaluation window; missing-data rules further determine retention. The 724 AWS winter-2026 fold retains 540 groups across 91 base times (Supplementary Note 10 and Supplementary Table 8). The 724-station Automatic Weather Station (AWS) network, 4.6–6.3 times denser in stations reporting all five variables, was evaluated on the same folds to test whether the results hold when the network changes; because network type changes along with density, it measures reproduction on a different network rather than the effect of density alone ('Replication across observing networks'). Regional replication ('Replication across regional observing networks') uses the NOAA GHCNh archive over ten regions on six continents, with the eight domestic folds read in UTC, stations selected before training by a fixed availability rule, inputs taken within 20 min before the base time and targets by nearest match within ±20 min, and the diurnal phase computed in local solar time. Quality control, missing-data handling and the full sampling design are given in Supplementary Note 12. The South Korean station observations and surface objective analysis were obtained from the Korea Meteorological Administration (KMA) API Hub and processed by the authors.

## Baselines and ablation

CLARA was compared with six external baselines and with one learned alternative given the same temporal-phase input. The statistical baselines are persistence, climatology and a per-lead ensemble model output statistics (EMOS) model; the learned baselines are iTransformer (variable-axis attention), the same model given the temporal-phase input of CLARA (iTransformer + phase) and graphAttn (station self-attention on the CLARA backbone with the advection-routing bias and the phase input removed); the physical baseline advects an observation-based objective analysis with the station wind. All except the physical baseline use the same base times, stations, leads and targets as CLARA and are recalibrated with the same lead-wise procedure. The physical baseline is deterministic and, because upwind points outside the analysis grid are treated as missing, is scorable on 81% of 96-station samples at 1 h and 43% at 48 h (79% and 31% on the 724-station network), so it serves as a physical reference rather than a matched competitor ('Overall forecast accuracy'). The original learned baselines have fewer parameters than CLARA (iTransformer 18,352, with phase input 18,640; graphAttn 27,595; CLARA 27,979).

Additional comparisons retrained iTransformer + phase at increased widths to obtain similar parameter counts: 27,616 versus 27,979 domestically, and 27,335 versus 27,136 regionally. Widths were chosen before inspecting these comparison results. Multi-year comparisons use the same fold and seed aggregation as the primary evaluation (Supplementary Notes 9 and 12).

Structural ablations retrain CLARA without one component on the same folds and seeds: without the advection-routing bias (no-advection), without the temporal-phase input (no-phase), with a single learned constant in place of the lead-conditioned coupling (constant-γ), and with the wind vector in the distance bias set to zero so that the upwind location coincides with the station (zero-wind; eight seeds on 96 ASOS and five on 724 AWS). Two further conditions on 96 ASOS locate where lead information enters: gateonly removes the lead channel from the token embedding, and constonly additionally replaces $\gamma(\tau)$ with a learned constant; both retain the target-hour phase and the lead-dependent upwind displacement ('Ablation and the routing mechanism', Supplementary Note 1). Per-lead specialists trained with the same number of optimization steps (three seeds, multi-year and 2026 folds) and a model whose Cholesky diagonal floor is lowered to 0.005 (three seeds, multi-year folds) serve as references for the unified design and for the floor. Each contribution is the fold-mean energy score of the condition minus that of the full model, relative to the full model (%), so a positive value means the component improves performance. Baseline implementations are given in Supplementary Note 9, the definition of every condition in Supplementary Note 7, and the results of the specialists and of the floor in Supplementary Notes 1 and 4.

## Evaluation metrics and statistical testing

Accuracy was assessed with proper scoring rules, calibration was diagnosed with the squared Mahalanobis distance and the probability integral transform, and predictive spread was assessed alongside calibration (Gneiting et al., 2007).

The primary metric for the multivariate joint predictive distribution is the energy score, a proper scoring rule that generalizes the CRPS to several variables; lower expected scores indicate better probabilistic predictions (Gneiting & Raftery, 2007). It was computed by Monte Carlo from two independent batches of 1,000 draws each from the predictive distribution. Forecast and observation enter the score in the standardized units in which the model is trained (Supplementary Note 8), so energy scores are dimensionless, whereas the per-variable RMSE and CRPS are reported in physical units. The standard-normal draws come from a counter-based generator keyed by base time, lead and station, so every model compared receives the same random numbers for a given sample (common random numbers), which limits the effect of Monte Carlo variation on the comparison, while different samples receive independent draws. Because each term is a sample mean whose expectation equals the corresponding term of the score, the estimator is unbiased, and its variance falls at order $1/n_s$ in the number of draws $n_s$. The definition of the estimator and its properties are given at the end of this section.

Per-variable deterministic accuracy was measured by RMSE, and the per-variable marginal predictive distribution by CRPS. For every score a lower value is better. CLARA minimizes a strictly proper

Gaussian log score in training and is evaluated with the strictly proper energy score. Although the scores differ, both expected scores are minimized by the true conditional distribution when the model is correctly specified.

Calibration of the joint predictive distribution was evaluated with the squared Mahalanobis distance $d^2$ defined above. If the mean and the covariance are correctly specified, $\mathbb{E}[d^2]$ equals the number of variables $V$, which makes the ideal value $\mathbb{E}[d^2] = 5$ here. In the Results and tables $\mathbb{E}[d^2]$ denotes the sample mean of $d^2$ over the samples scored at a lead. The multivariate PIT was also computed, by applying the chi-squared distribution function with five degrees of freedom to $d^2$ and checking how close the result is to uniform. A multivariate diagnostic alone cannot reveal every per-variable misspecification, so the PIT of each variable was examined as well. The calibration diagnostics are detailed in Supplementary Note 13.

The contribution of the dependence structure learned by the joint output layer was evaluated post hoc on the stored forecasts of the 96-station multi-year models (all eight seeds), with no retraining ('Joint uncertainty and recalibration'; 'Derived quantities and compound events'). With each marginal variance held fixed, the learned correlation matrix was replaced either by the identity (conditional independence) or by a fixed empirical correlation matrix estimated per lead from standardized residuals on the training split, and the same forecasts were rescored with the joint negative log-likelihood, the energy score and the variogram score of order 0.5. The variogram score is comparatively sensitive to misspecification of the dependence between variables, whereas the energy score can be insensitive to it (Scheuerer & Hamill, 2015).

Derived-quantity distributions and compound-event probabilities were evaluated under all three correlation structures, analytically for the pressure difference and by Monte Carlo otherwise (Supplementary Note 5). The targets were the difference between sea-level and station pressure ps − pa, the discomfort index, wind chill (on the rows where it is defined, temperature at most 10 °C and wind speed above 3 mph) and a compound event, defined as temperature above the 90th percentile of the training split together with humidity below the 10th percentile. This comparison measures what accounting for the covariance between variables does to estimates of derived quantities and of compound-risk probabilities.

The synthetic control used the 96 station locations and three variables. Base-time states $u_i$ were drawn from a spatially correlated Gaussian field (exponential covariance with a 150 km scale, correlation 0.3 between variables), and winds from a sample-wide direction and a speed of 2–12 m/s with station-level direction noise. Each target was $y_i = a(\tau)\, u_i + \omega(\tau) \sum_j \rho_{ij}\, u_j + \varepsilon$, where $\rho_{ij}$ normalizes $\exp(-d(\mathrm{up}_i(\tau), p_j)/60\,\mathrm{km})$ over stations, $\mathrm{up}_i(\tau)$ is the spherical upwind location used by CLARA ('Lead-conditioned advection-routing attention') and $\omega(\tau)$ is 0.2, 0.6, 0.8, 0.5, 0.3 and 0.1 at the six leads; a control data set sets $\omega \equiv 0$. CLARA was trained with the optimization settings of the main experiments and three seeds per data set; the synthetic inputs and generating parameters are specified in Supplementary Note 3. The model effect of an intervention is the change of the predictive mean at a target per unit perturbation of one variable at the five non-self stations $J_i$ with the largest true kernel weight, and the true effect is $\omega(\tau) \sum_{j \in J_i} \rho_{ij}$; both were computed on the same samples and targets (Supplementary Note 3).

To test whether relative logit size determines forecast influence, the distance-bias coefficient and content logits of the stored 96-ASOS models were rescaled at inference without retraining. Equal multipliers preserve their relative size. We compared energy scores with and without the distance term on identical evaluation rows (Fig. 5f; Supplementary Note 3).

Because there are only six multi-year folds and their training windows overlap, differences between folds were not treated as independent replicates and no formal significance test was performed. Results are reported as fold means of seed means, each of which averages the scored rows with equal weight, together with the range across folds and the number of folds in which the direction holds; root-mean-square errors are formed from the fold-mean squared error. Unless otherwise stated, the primary learned-model comparisons used eight seeds per fold on the 96 ASOS stations, five on the 724 AWS stations and three per region on GHCNh; the per-lead specialists and the covariance-floor diagnostic used three seeds.

The two 2026 seasons were held out: they were not used for model development or for choosing any setting. For each of the two 2026 folds, models were trained from scratch on that fold's preceding observations using the fixed architecture, training settings and calibration procedure, without transferring weights from the multi-year folds. The capacity-controlled and regional learned-model comparisons were additional retrained evaluations, outside the pre-specified primary comparisons (Supplementary Note 10). CLARA is described as better at every lead in the two later evaluation seasons only if its seed-mean energy score is lower than that of each of the five pre-specified baselines (persistence, climatology, EMOS, iTransformer and graphAttn) in both 2026 seasons at all six leads. iTransformer + phase is not part of this rule.

## Unbiasedness of the energy-score estimator

For a predictive law $F$ and observation $y$, $\mathrm{ES}(F, y) = \mathbb{E}\|X - y\| - 1/2\,\mathbb{E}\|X - X'\|$ with $X, X' \sim F$ independent. With two independent batches $\{X_k\}_{k=1}^{n_s}$ and $\{X'_k\}_{k=1}^{n_s}$ drawn from $F$ ($n_s = 1{,}000$ throughout), the estimator

$$\widehat{\mathrm{ES}} = \frac{1}{n_s}\sum_{k=1}^{n_s}\|X_k - y\| - \frac{1}{2n_s}\sum_{k=1}^{n_s}\|X_k - X'_k\| \quad (10)$$

is unbiased: each $\|X_k - y\|$ has expectation $\mathbb{E}\|X - y\|$, and by independence of the two batches each pair $(X_k, X'_k)$ is an independent draw from $F \otimes F$, so $\mathbb{E}\|X_k - X'_k\| = \mathbb{E}\|X - X'\|$. Each term averages $n_s$ i.i.d. summands with finite second moments, so the variance of $\widehat{\mathrm{ES}}$ is of order $1/n_s$; the two terms are correlated through the shared $X_k$, which does not change that order. Pacchiardi et al. (2024) give an unbiased estimator of the same score that averages the second term over all distinct pairs within one batch; the paired form of equation (10) needs $n_s$ distance evaluations for the second term rather than one for each of the $n_s(n_s - 1)/2$ distinct pairs.

## Implementation and training settings

The station token is formed by a linear layer of latent dimension $d_{\mathrm{model}} = 64$ with GELU; spatial combination uses one multi-head self-attention block with four heads, whose output passes through a sigmoid-gated residual, layer normalization and a feed-forward network into a mean head and a Cholesky head whose diagonal is bounded below by adding 0.05 to a softplus. The multilayer perceptron producing $\gamma(\tau)$ is 1–16–1, and the lead enters it as $\tau_{\mathrm{norm}} = (\ln\tau - \ln 360)/0.9$ with $\tau$ in minutes. The model, with elevation as its only static feature, has 27,979 trainable parameters. Training ran for 40 epochs with Adam at a fixed learning rate of $1.5 \times 10^{-3}$, with no schedule, weight decay, dropout or early stopping; these settings were fixed across the reported folds ('Observations and preprocessing'). A forward pass producing the joint distribution for one base time and lead at the 96 stations took 0.43 ms on one thread of an Intel Core i9-12900 (PyTorch 2.5.1; the experiments themselves used PyTorch 1.13.1), and a complete 40-epoch training run took 2.1 min on one thread and 1.7 min on an NVIDIA GeForce RTX 3080 GPU, with peak host memory below 700 MB in the CPU runs. Architecture and training settings are given in Supplementary Note 8, and the cost measurements in Supplementary Note 14 and Supplementary Table 9.

### Limits of the evaluation design

The evaluation is conditional on stations reporting all variables of the joint distribution; robustness to the removal of context stations was tested (Extended Data Fig. 2), but per-variable sensor dropout and extreme observation gaps were not. Further limits and their measurements are set out in Supplementary Note 1.

## Data availability

The South Korean station observations (ASOS and AWS) and surface objective analysis are available from the Korea Meteorological Administration (KMA) API Hub (https://apihub.kma.go.kr; registration required). The NOAA GHCNh station-year files used for the ten-region evaluation are publicly available from NOAA NCEI (https://www.ncei.noaa.gov/oa/global-historical-climatology-network/hourly/access/by-year/). The HRRR forecasts are available from the NOAA Open Data Dissemination archive (https://noaa-hrrr-bdp-pds.s3.amazonaws.com), and the ECMWF ensemble forecasts from the TIGGE archive through the ECMWF Data Store (https://ecds.ecmwf.int; registration required).

## Extended Data

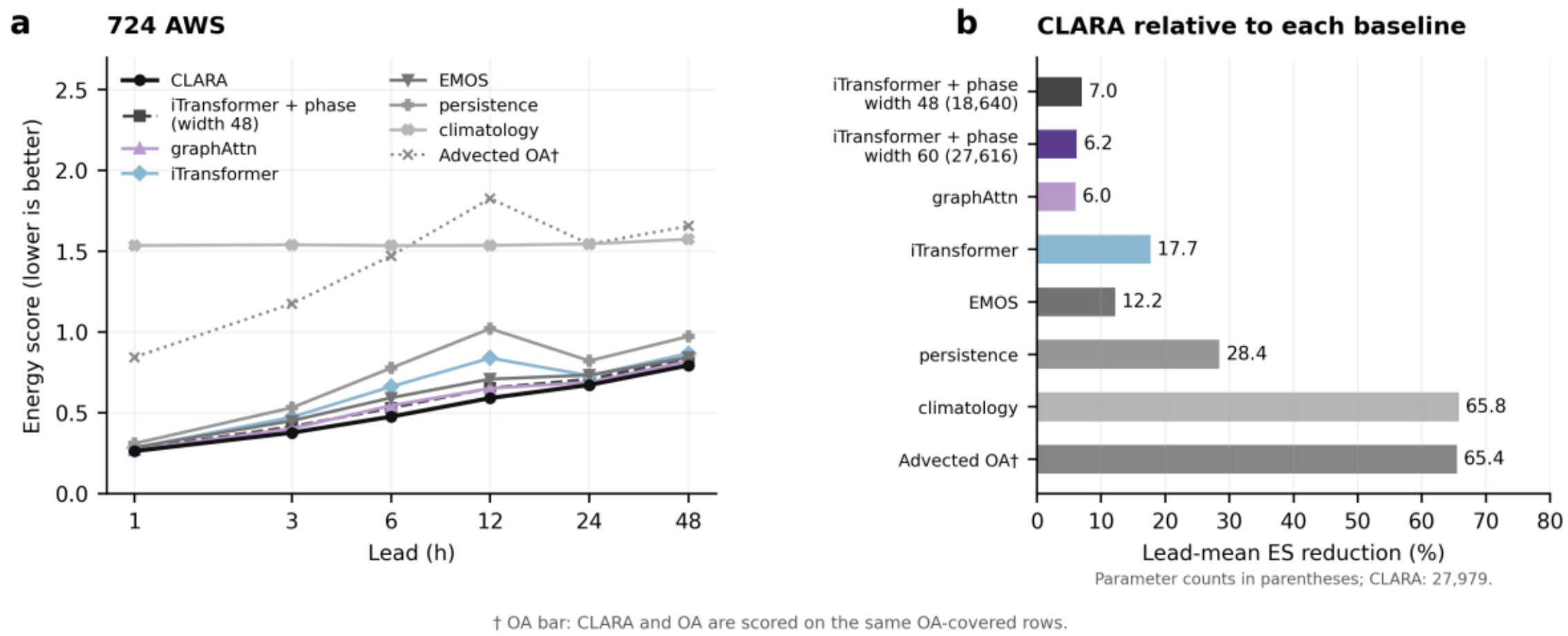

*Extended Data Fig. 1. Forecast comparison on the 724-station AWS network (six multi-year folds, five seeds for learned models).* ***a****, Fold-mean energy scores by lead. Probabilistic models use their issued distributions on the full evaluation set; the deterministic advected objective analysis (OA; †) is evaluated only on the rows it covers. The CLARA curve uses the full evaluation set.* ***b****, Lead-mean energy-score reduction of CLARA relative to each baseline, 100 ($S_{ref} - S_{CLARA}$)/$S_{ref}$, where each S is the mean energy score across the six leads. The OA comparison uses the same OA-covered rows for both methods. iTransformer + phase receives the same temporal-phase input as CLARA; graphAttn and iTransformer do not receive temporal-phase input. Panel a uses width 48; panel b also includes width 60, retrained with the same training, recalibration and evaluation protocol to give a parameter count similar to CLARA (27,616 versus 27,979; Supplementary Note 9). Parameter counts are shown in parentheses for the two widths. Matched-row scores and OA coverage are shown in Supplementary Fig. 2.*

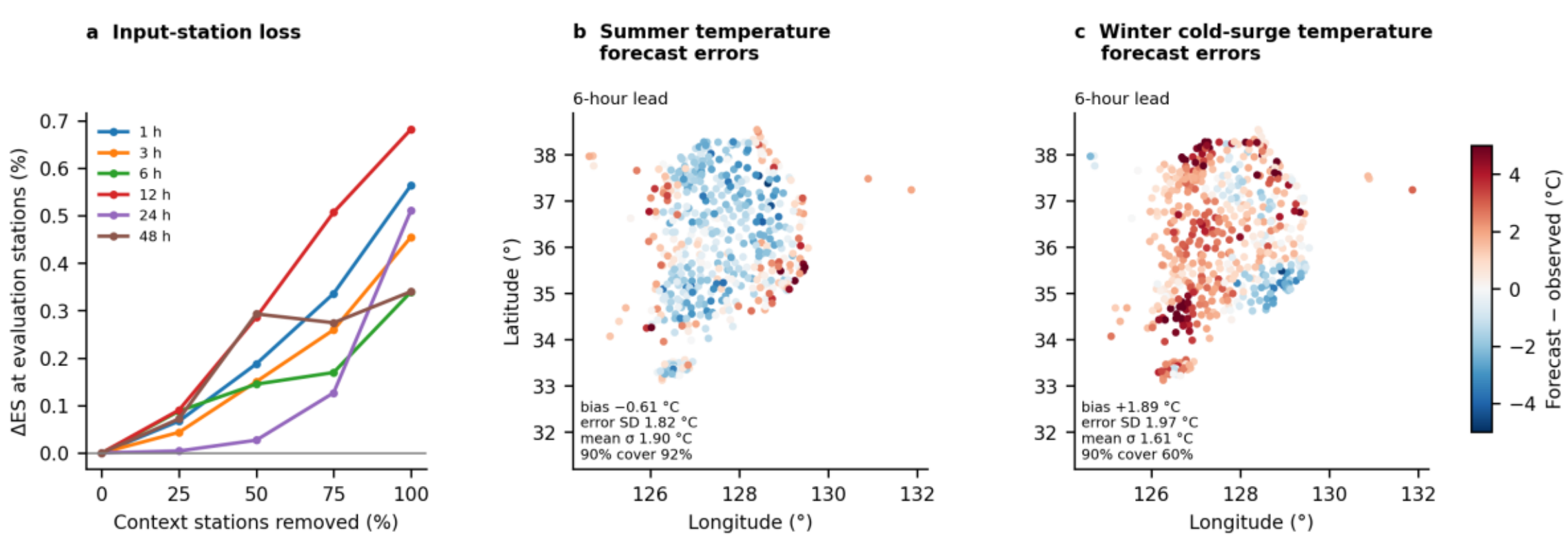

*Extended Data Fig. 2. Robustness to the loss of input stations and case forecasts (724 AWS, stored multi-year models, inference only). a, Change in the energy score at fixed evaluation stations, which stay in the input, as 0–100% of the remaining context stations are removed (three seeds, six folds). b, c, 6 h temperature error (forecast mean − observation, seed 0) for a summer case with the highest domain-mean target temperature in summer 2024 and a winter case with the strongest domain-mean 6 h cooling in winter 2024 (selection rules fixed in advance); inset statistics average five seeds.*

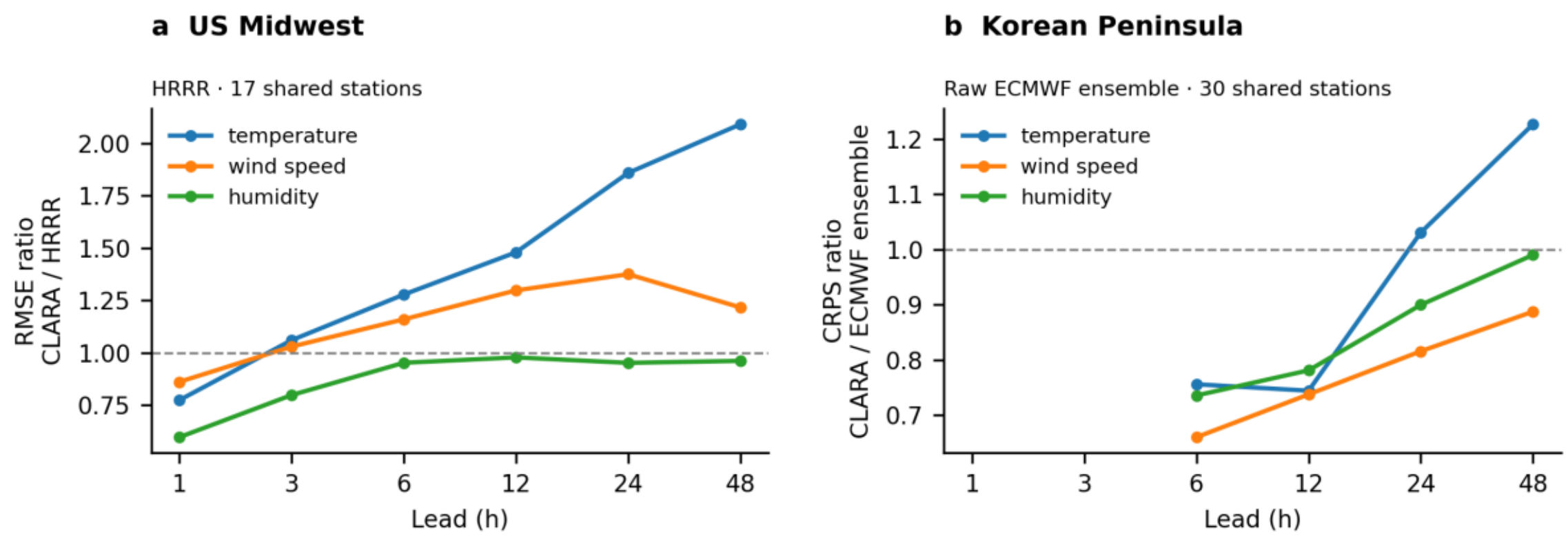

*Extended Data Fig. 3. Operational NWP references on shared stations, a diagnostic comparison (neither reference is post-processed to stations; CLARA is trained and recalibrated on them). a, US Midwest: RMSE of CLARA relative to HRRR (17 shared stations, two folds). b, Korean Peninsula: CRPS of the issued CLARA distribution relative to the raw ECMWF ensemble (51 members; 30 shared stations, six folds, 00 and 12 UTC; leads of 6–48 h). Values below one favour CLARA.*

Extended Data Table 1. System-level comparison of representative intelligent forecasting approaches with CLARA. Cells summarize the cited papers as characterized under 'Related approaches' in the Methods and in Supplementary Note 6; dashes mark aspects not applicable or not characterized there.

| Approach | Input data | Output and resolution | Lead range | Probabilistic output | Lead-time handling | NWP / reanalysis in the loop |
|---|---|---|---|---|---|---|
| MetNet-3 (Andrychowicz et al., 2023) | Multi-source inputs including assimilated atmospheric analyses | Gridded; surface variables at 4 km · 5 min (precipitation 1 km · 2 min) | to 24 h | Per-variable marginal (categorical) | Lead-conditioned, single pass | Assimilated fields as input |
| Aardvark Weather (Allen et al., 2025) | Raw multi-source observations | Gridded and station; two surface variables at stations, deterministic | to 10 d | None (deterministic) | Daily autoregressive roll-out | Processor pretrained on reanalysis |
| Probabilistic Aardvark (Almeida et al., 2026) | Raw multi-source observations | Gridded (24 variables, 1.5°) and station (2-m temperature, 10-m wind) | 1–10 d | Ensemble (input noise × MC dropout, 49 members) | Daily autoregressive roll-out | ERA5 training targets and gridded verification |
| AIFS-DOP (Pinnington et al., 2026) | Observations only (training and initialization) | Global grid ≈100 km, 6-h slices, deterministic | to 10 d | None (squared-error training) | Iterative roll-out | None |
| EMOS / neural post-processing (Gneiting et al., 2005; Rasp & Lerch, 2018; Baran & Möller, 2017) | NWP or ensemble forecasts | Station; mostly univariate predictive distribution | per lead | Mostly univariate (bivariate EMOS: joint wind speed and temperature) | Separate model per lead | NWP input required |
| Cholesky distributional regression (Muschinski et al., 2024) | NWP ensemble | Station; joint temperature distribution across leads | — | Multivariate joint (Cholesky) | Joint across leads | NWP input required |
| AGSTAN (Liang et al., 2024) | Gridded fields | Grid; ENSO prediction, deterministic | seasonal–interannual | None | Lead-varying attention | — |
| Corrformer (Wu et al., 2023) | Station observations only (hourly temperature and wind speed, past 48 h) | Station; hourly point series for 3,850 worldwide (or 34,040 Chinese) stations, deterministic | to 24 h | None (MSE training) | Direct multi-step output (24 hourly steps in one pass) | None (GFS and ERA5 used only as baselines) |
| CLARA (this work) | Station observations only | Station; five surface variables jointly | 1–48 h | Calibrated multivariate joint (μ, Cholesky Σ) | Single pass per lead; learned γ(τ), recalibrated s(τ) | None at any stage |

*Note. The comparison concerns the operating point (inputs, outputs, probabilistic form, lead handling), not skill: the listed systems target different domains and horizons and are not directly comparable on a common score.*

Extended Data Table 2. Multi-year nationwide comparison of the issued joint predictive distribution by lead time: energy score (fold mean) of CLARA and of the comparison models, with the improvement of CLARA in parentheses. 96 ASOS stations, 2021–2025, six rolling-origin folds (winter and summer test seasons 2023–2025), eight seeds for the learned models.

| Lead | CLARA | iTransformer + phase | Persistence | Climatology | EMOS | Advected OA |
|---|---|---|---|---|---|---|
| 1 h | 0.279 | 0.286 (+2.5%) | 0.330 (+15.5%) | 1.542 (+81.9%) | 0.300 (+6.9%) | 0.906 (+70.4%) |
| 3 h | 0.393 | 0.422 (+6.7%) | 0.555 (+29.2%) | 1.548 (+74.6%) | 0.462 (+14.8%) | 1.415 (+73.4%) |
| 6 h | 0.492 | 0.533 (+7.6%) | 0.809 (+39.2%) | 1.540 (+68.0%) | 0.595 (+17.2%) | 1.850 (+74.5%) |
| 12 h | 0.605 | 0.650 (+6.9%) | 1.064 (+43.1%) | 1.540 (+60.7%) | 0.709 (+14.7%) | 2.319 (+75.0%) |
| 24 h | 0.693 | 0.713 (+2.8%) | 0.844 (+17.8%) | 1.549 (+55.3%) | 0.750 (+7.6%) | 2.154 (+69.6%) |
| 48 h | 0.817 | 0.845 (+3.3%) | 1.005 (+18.7%) | 1.583 (+48.4%) | 0.870 (+6.1%) | 2.332 (+67.4%) |

*Note. ES ↓, lower is better; improvement $I = 100\,(S_{ref} - S_{CLARA})/S_{ref}$, positive when CLARA performs better. CLARA's fold-mean energy score is lower than that of every listed model at all six evaluated leads and lower in 143 of the 144 fold–lead cells against iTransformer + phase and the three statistical baselines. In the two 2026 held-out seasons, CLARA's seed-mean energy score is lower than that of the five pre-specified baselines (persistence, climatology, EMOS, iTransformer and graphAttn) in all 60 season–lead–baseline cells. iTransformer + phase receives the same temporal-phase input as CLARA. Learned models without phase input are reported in Supplementary Table 2. The advected OA is a deterministic physical baseline scored only on the rows whose upwind point lies inside the analysis grid (coverage 0.43–0.83 by lead); its column compares CLARA on those same rows.*

Extended Data Table 3. Calibration before and after recalibration on the 96-station multi-year folds (six folds, eight seeds): the fitted lead-conditioned scale s(τ) (last 15% of each training window, excluded from training), the mean squared Mahalanobis distance $E[d^2]$ (ideal V = 5) of the raw distribution, with a single global scale and with s(τ), the deviation $D = |E[d^2] - 5|$ before and after s(τ), and the energy score before and after.

| Lead | s(τ) | $E[d^2]$ raw | $E[d^2]$ global scale | $E[d^2]$ with s(τ) | $D = \lvert E[d^2] - 5 \rvert$ raw → with s(τ) | Folds with lower D | Energy score raw → with s(τ) |
|---|---|---|---|---|---|---|---|
| 1 h | 0.78 | 3.11 | 3.05 | 5.12 | 1.89 → 0.52 | 6/6 | 0.2802 → 0.2792 |
| 3 h | 0.88 | 3.83 | 3.75 | 4.93 | 1.17 → 0.43 | 6/6 | 0.3936 → 0.3933 |
| 6 h | 0.97 | 4.46 | 4.38 | 4.79 | 0.63 → 0.66 | 4/6 | 0.4924 → 0.4923 |
| 12 h | 1.09 | 5.34 | 5.25 | 4.58 | 0.64 → 0.90 | 2/6 | 0.6033 → 0.6050 |
| 24 h | 1.15 | 5.62 | 5.57 | 4.48 | 0.88 → 1.48 | 2/6 | 0.6874 → 0.6930 |
| 48 h | 1.17 | 5.62 | 5.54 | 4.23 | 0.91 → 1.16 | 2/6 | 0.8108 → 0.8171 |

*Note. D is computed per run and averaged over seeds and then folds, so it is not $|\text{mean } E[d^2] - 5|$. ES ↓, lower is better. The global scale is close to one (fold mean 1.02) and does not reduce D at any lead. The same recalibration applied to every baseline preserves the ranking (Supplementary Table 4).*

Extended Data Table 4. Derived quantities under three correlation structures (96 ASOS multi-year folds, eight seeds, issued distribution): the mean predictive standard deviation of the pressure difference ps − pa when the learned correlations are replaced by conditional independence (diag) or by a fixed empirical correlation (fixed), relative to the learned correlations; $E[z^2]$ of the standardized ps − pa under the learned correlations (ideal 1); the change in PIT variance ratio for the discomfort index and wind chill (on the rows where wind chill is defined) under conditional independence;

and the relative change in the Brier score of the compound hot–dry event (temperature above the training-split 90th percentile and humidity below the 10th).

| Lead | ps − pa predictive SD diag / learned | ps − pa predictive SD fixed / learned | ps − pa E[z²] learned (ideal 1) | Discomfort index \|Δ PIT var. ratio\| | Wind chill \|Δ PIT var. ratio\| | Compound event ΔBrier diag / fixed (%) |
|---|---|---|---|---|---|---|
| 1 h | 2.7 | 0.52 | 0.10 | 0.051 | 0.058 | +0.7 / +0.1 |
| 3 h | 3.0 | 0.48 | 0.10 | 0.071 | 0.048 | +0.5 / 0.0 |
| 6 h | 3.4 | 0.50 | 0.11 | 0.069 | 0.032 | −3.2 / −1.2 |
| 12 h | 4.7 | 0.54 | 0.12 | 0.078 | 0.014 | −5.0 / −2.1 |
| 24 h | 7.4 | 0.55 | 0.12 | 0.093 | 0.009 | −4.2 / −3.3 |
| 48 h | 10.2 | 0.66 | 0.15 | 0.111 | 0.004 | −4.7 / −4.0 |

*Note. Positive Brier changes mean the alternative is worse than the learned correlations. E[z²] below one means that the predictive distribution of ps − pa is too wide ('Derived quantities and compound events' in the main text).*

Extended Data Table 5. Structural ablations on the same backbone, multi-year folds: change in the energy score of the issued distribution when a component is removed and the model retrained (ablation − full, relative to full; positive = the component improves performance), with the number of folds in which the change is positive, and the learned γ(τ) of CLARA. 96 ASOS (eight seeds) and 724 AWS (five seeds), including zero-wind on both networks, six rolling-origin folds.

| Lead | no-phase 96 ASOS | no-phase 724 AWS | no-advection 96 ASOS | no-advection 724 AWS | constant-γ 96 ASOS | constant-γ 724 AWS | zero-wind 96 ASOS | zero-wind 724 AWS | CLARA γ(τ) 96 / 724 |
|---|---|---|---|---|---|---|---|---|---|
| 1 h | +0.85% [5/6] | +0.89% [4/6] | +0.78% [6/6] | +2.19% [6/6] | +0.36% [5/6] | +1.33% [6/6] | −0.25% [2/6] | +0.14% [3/6] | 1.61 / 3.22 |
| 3 h | +5.30% [6/6] | +4.71% [6/6] | +1.28% [6/6] | +2.82% [6/6] | +0.46% [5/6] | +1.37% [6/6] | −0.26% [2/6] | +0.13% [5/6] | 1.03 / 1.81 |
| 6 h | +15.23% [6/6] | +13.19% [6/6] | +0.87% [6/6] | +2.00% [6/6] | −0.07% [2/6] | +0.78% [6/6] | −0.28% [2/6] | −0.10% [2/6] | 0.68 / 1.04 |
| 12 h | +11.41% [6/6] | +9.26% [6/6] | +0.73% [6/6] | +1.05% [5/6] | −0.07% [4/6] | +0.27% [5/6] | −0.10% [3/6] | −0.08% [3/6] | 0.40 / 0.47 |
| 24 h | +2.20% [6/6] | +2.32% [6/6] | +0.24% [4/6] | +0.51% [4/6] | +0.03% [1/6] | +0.12% [4/6] | +0.02% [5/6] | +0.02% [4/6] | 0.24 / 0.18 |
| 48 h | +3.50% [5/6] | +2.93% [5/6] | +0.25% [5/6] | +0.56% [5/6] | +0.18% [5/6] | +0.18% [3/6] | +0.31% [4/6] | +0.11% [4/6] | 0.15 / 0.06 |

*Note. Fold values are seed means; entries are fold means. No intervals are given because the six folds share training windows; the number of folds with a positive change is reported instead. No-phase removes the temporal-phase input; no-advection removes the advection-routing bias; constant-γ replaces γ(τ) with a single learned constant; zero-wind sets the wind vector in the distance bias to zero, so the upwind location coincides with the station. γ(τ) is the fold mean of the seed means read from the stored weights.*

Extended Data Table 6. Regional replication on GHCNh: improvement (%) of the issued energy score of CLARA over persistence and over climatology by region and lead (six multi-year folds read in UTC, three seeds; up to 50 stations per region; temperature, wind speed and humidity).

| Region | 1 h vs persistence / climatology | 3 h vs persistence / climatology | 6 h vs persistence / climatology | 12 h vs persistence / climatology | 24 h vs persistence / climatology | 48 h vs persistence / climatology |
|---|---|---|---|---|---|---|
| Korean Peninsula | 14.8 / 76.4 | 28.7 / 67.4 | 36.8 / 59.6 | 38.9 / 51.9 | 16.5 / 45.7 | 18.2 / 41.3 |
| Central Europe | 17.3 / 75.3 | 32.4 / 65.9 | 40.6 / 58.0 | 45.4 / 49.2 | 17.4 / 41.6 | 17.5 / 34.4 |
| US Midwest | 17.5 / 72.3 | 33.6 / 63.0 | 41.8 / 55.2 | 46.8 / 46.1 | 21.8 / 38.5 | 22.2 / 32.0 |
| Scandinavia | 11.7 / 78.0 | 23.2 / 68.1 | 32.1 / 59.8 | 36.3 / 50.9 | 19.0 / 44.0 | 19.9 / 37.7 |
| Southeast Asia | 24.6 / 47.6 | 40.0 / 46.7 | 55.4 / 41.0 | 59.6 / 36.9 | 17.1 / 38.7 | 19.1 / 33.9 |
| Central Brazil | 21.3 / 57.7 | 41.9 / 47.7 | 53.7 / 39.8 | 59.0 / 35.4 | 14.1 / 29.0 | 19.3 / 30.2 |
| West Africa | 17.9 / 63.3 | 40.3 / 52.2 | 54.7 / 48.4 | 61.0 / 44.4 | 17.9 / 46.7 | 18.8 / 43.6 |
| East Africa | 20.6 / 59.0 | 40.3 / 50.3 | 53.1 / 43.9 | 58.7 / 40.8 | 16.7 / 43.2 | 18.6 / 41.0 |
| Southeastern Australia | 15.3 / 72.4 | 32.7 / 60.9 | 43.5 / 50.5 | 49.7 / 42.7 | 17.4 / 40.2 | 20.6 / 33.4 |
| Middle East | 15.0 / 72.7 | 30.7 / 64.3 | 43.7 / 58.2 | 48.8 / 53.5 | 16.6 / 52.3 | 20.7 / 47.6 |

*Note. $I = 100 (S_{ref} - S_{CLARA})/S_{ref}$, positive when CLARA performs better. The joint distribution is restricted to three variables because station and sea-level pressure are reported at base hours by fewer than half of the qualifying stations in several regions. Central Brazil, East Africa and southeastern Australia have 36, 13 and 15 qualifying stations. In both 2026 held-out seasons CLARA also outperforms persistence in every region–lead combination.*

Extended Data Table 7. Datasets and their experimental roles. Scored stations are those with all variables of the joint distribution valid at base and target time (fold range of the per-sample mean). Fold-level details are in Supplementary Table 8.

| Dataset | Period | Stations (scored per sample) | Protocol | Seeds | Role |
|---|---|---|---|---|---|
| ASOS 96 (nationwide) | 2021–2026 | 96 (95–96) | 6 multi-year folds (JJA/DJF 2023–2025) + 2 held-out folds (Dec 2025–Feb 2026, Jun–Aug 2026); base times every 4 days at 00, 06, 12, 18 KST | 8 | Main benchmark, calibration, ablations, dependence |
| AWS 724 (nationwide) | 2021–2026 | 724 nominal (436–601) | Same 8 folds | 5 | Replication on a denser network |
| GHCNh (10 regions) | 2021–2026 | 13–50 per region | Same 8 folds read in UTC; stations selected before training by availability | 3 | Regional replication |

*Note. Density and network type (ASOS vs AWS) change together between the two domestic networks. The GHCNh joint distribution comprises temperature, wind speed and humidity.*

## Acknowledgements

This research was supported by Global - Learning & Academic research institution for Master's·PhD students, and Postdocs (G-LAMP) Program of the National Research Foundation of Korea (NRF) grant funded by the Ministry of Education (No. RS-2026-25610465).

# Supplementary Information — Learning joint probabilistic weather forecasts from station observations alone

## Contents

Supplementary Notes 1–5 report additional results and the dependence-analysis methods: further evaluations and diagnostic contrasts, a reference comparison with Corrformer, the measurements behind the reading of the learned coefficient, the effect of the covariance floor and the methods of the dependence analysis. Supplementary Note 6 reviews related approaches. Supplementary Notes 7–14 give the ablation and decomposition variants, the network and training details, the baselines, the validation protocol, the assumptions and proofs, the observations, the calibration diagnostics and the computational cost.

*Notation follows the main text (equations (1)–(10) and Lemma 1); equations (S1)–(S13) are those of this document.*

**Supplementary Notes**

**Supplementary Figures**

**Supplementary Tables**

**Supplementary References**

## Supplementary Note 1. Additional evaluations and diagnostic contrasts

This note reports measurements that qualify the main results. Unless stated otherwise, values are fold means of seed means on the six multi-year folds of the 96-station network, and a positive change means that the component improves performance.

**Lead-conditioned coupling.** Replacing $\gamma(\tau)$ by a single learned constant (constant-γ) raises the energy score by 0.36% and 0.46% at 1 and 3 h, in five of the six folds, and changes it by −0.07% to +0.18% at 6–48 h. On the 724-station network the rise is 1.33% and 1.37% at 1 and 3 h and 0.78% at 6 h, in all six folds, and at most 0.27% at 12–48 h (Extended Data Table 5).

**Where the lead enters.** Holding the lead channel of the token embedding at zero, so that the lead enters only through $\gamma(\tau)$ (gateonly), raises the energy score relative to the full model by 5.5% at 1 h and by 0.9–1.8% at 3–48 h. Removing the lead from both the token embedding and the coupling (constonly) raises it by 3.3% at 1 h and by 0.4–1.5% at 3–48 h. Keeping the lead in $\gamma(\tau)$ alone is therefore worse than removing it from both pathways at every lead, in four or five of the six folds. Both variants retain the target-hour phase and the lead-proportional upwind displacement (Supplementary Note 7).

**Per-lead specialists.** Six specialists, each trained on a single lead with the same number of optimization steps as the unified model (three seeds), were scored on the same rows as the unified model. The specialists are better at 3 h in all six folds (by 3.1%) and at 6 h in four (by 0.7% in the fold mean), and worse at 12, 24 and 48 h in every fold (by 5.0%, 16.0% and 19.6%). At 1 h, specialists score better in 11 of the 18 runs and worse in seven. In four of the latter runs, their energy score is higher by 0.18–0.48. Across all runs, the fold-mean disadvantage is 20.4%. Relative differences divide the fold-mean paired score difference by the fold-mean CLARA score from the same three seeds. In the two 2026 seasons, specialists are better at 1 and 3 h (by 2.0% and 2.1%) and worse at 12–48 h (by 4.9–16.3%); at 6 h their score is 0.2% higher, with opposite directions in the two folds. The unified model thus has lower energy scores than the specialists at 12–48 h in every multi-year fold, and higher ones at 3 h in the multi-year folds and at 1 and 3 h in the 2026 seasons. Together, the six specialists require about six times as many optimization steps as the unified model.

**Wind in the distance bias.** The zero-wind retraining on both networks is described in the main text ('Ablation and the routing mechanism'). Here the scored rows are further divided by the base-time wind speed at the scored station into five strata, bounded by the 25th, 50th, 75th and 90th percentiles of each network's scored rows (0.7, 1.5, 2.5 and 3.8 m s$^{-1}$ on 96 ASOS), and the energy-score difference between the zero-wind and the full model is averaged within each run and then over seeds and folds. On 96 ASOS, the fold-mean difference lies between −0.0038 and +0.0038 in every stratum and lead, and in every stratum–lead cell it is positive in some folds and negative in others. In the stratum at or above the 90th percentile it is −0.0038 to −0.0004 at all six leads, positive in one to four of the six folds. On 724 AWS, the stratum boundaries are 0.7, 1.6, 2.8 and 4.4 m $s^{-1}$. Across strata and leads, the fold-mean difference ranges from −0.0039 to +0.0022. In its strongest-wind decile, the range is −0.0039 to +0.0010, positive in one to five of the six folds. Every strongest-wind stratum–lead cell has both positive and negative fold differences.

**Dependence on the 724-station network.** On the 724-station network (five seeds) the fixed empirical correlation does not share the advantage it has on the 96-station network: its log-likelihood differs from that of the learned correlations by −0.06 to +0.09 nats at 1–24 h and is worse at 48 h in all six folds (by 0.30 nats), and it improves the CRPS of the pressure difference by at most 12% (25–32% at 1–24 h on the 96-station network). Removing the correlations worsens the log-likelihood by 1.4–2.8 nats per station and the energy score by 0.15–0.65%, in all six folds at every lead, as on the 96-station network.

**Evaluation in the two 2026 seasons.** For each held-out 2026 season, models were trained from scratch on the preceding observations with the same architecture, training settings and recalibration procedure, and then evaluated. CLARA has a lower issued energy score than each of the five pre-specified baselines in all 60 season–lead–baseline comparisons on each domestic network. Against iTransformer + phase, an additional comparison, CLARA has a lower score in all 12 season–lead comparisons on each network, with lead-mean reductions of 6.1% on 96 ASOS and 8.5% on 724 AWS. Against the wider, similarly sized comparator, the reductions are 5.8% and 7.8% (additional retrained comparison; Supplementary Note 9).

## Supplementary Note 2. Reference comparison with Corrformer

Corrformer (Wu et al., 2023) is the published deep model closest to this study in that it forecasts station observations directly from station observations. It was run as a reference on the six multi-year folds of the 96-station network with its public implementation (commit e3e0c6d), whose model code was used unchanged; the 96 stations were arranged in its two-level hierarchy as eight latitude bands of twelve stations, ordered by longitude within each band. The comparison was added after the 2026 folds had been scored, so it uses the multi-year folds only, and none of its settings was chosen on evaluation scores.

The input window ends at the base time and covers the preceding 48 hourly observations of all 96 stations and five variables, standardized with the statistics of CLARA's training split. Missing values were filled forward in time with the last available value, and with zero (the training mean in standardized units) before the first observation, so no input uses observations after its window; the training and validation losses count only observed targets, and only observed targets were scored. The model emits the 48 hourly steps after the base time in one pass (label length 24 h), from which the six leads are read. It was trained with the mean squared error and Adam at a learning rate of $10^{-4}$, in batches of four windows taken every 6 h, using only windows whose base time and targets lie inside the training split, for at most ten epochs with early stopping on the last 15% of those windows (patience three); training windows whose targets reach the validation windows were excluded, and the calibration split was used only for $s(\tau)$. The validation error reached its minimum at epoch 4–9 in 17 runs and at the tenth and last epoch in one, and the weights of the epoch with the lowest validation error were used. With a width of 128 in this reference configuration (the public default is 512) the model has 23,949,071 parameters, about 860 times as many as CLARA.

Because Corrformer issues point forecasts, its predictive distribution was formed as for persistence and EMOS, from the per-lead residual covariance of its predictions on the training split, and recalibrated with $s(\tau)$ fitted on the calibration split. Three seeds were run for each fold, and energy scores were computed on the same rows as CLARA with the same common random numbers.

The issued energy score of CLARA is lower than that of Corrformer at every lead in every fold (Supplementary Table 3). The improvement is 35.3% at 1 h, 21.6% at 3 h, 13.4% at 6 h and 7.2–9.6% at 12–48 h. Corrformer is worse than persistence at 1 h in all six folds, better at 3 h in four and better at 6–48 h in all six. The same ranking holds for the predictive mean: on the same rows the root-mean-square error in standardized units is 0.310 for Corrformer against 0.219 for CLARA at 1 h and 0.632 against 0.592 at 48 h, and CLARA's is lower in all six folds at every lead except 12 h (five of six), so the gap is not due to the construction of Corrformer's predictive distribution alone. The reference differs from CLARA in its objective (a point forecast with a fixed residual covariance rather than a learned joint distribution), in capacity and in its input window, and its hyperparameters were not tuned for this network, so it is not a matched comparison of architectures.

## Supplementary Note 3. Measurements for reading the learned coefficient

**Content scale and relative size of the distance term.** The quantities of Lemma 1 were measured on the stored multi-year models without retraining. For each query row and head the content logit $e_{ij}$ — the query–key score plus the salience offset — and the distance term $\gamma(\tau)A_{ij}$, with $A_{ij} = d_{ij}^{\mathrm{up}}/100$ as in main-text equations (2) and (4), were formed. With $\alpha_{ij} = \mathrm{softmax}_j(e_{ij} - \gamma A_{ij})$ the attention weights and $\alpha_{ij}^0 = \mathrm{softmax}_j(e_{ij})$ the weights with the distance term removed, the content scale, the size of the distance term relative to the content path, the effect on the weights and the pull towards the upwind position are

$$\lambda_i = \mathrm{SD}_j(e_{ij}), \qquad \frac{\mathrm{SD}_j(\gamma A_{ij})}{\lambda_i}, \qquad \mathrm{KL}(\alpha_i \parallel \alpha_i^0) = \sum_j \alpha_{ij} \ln \frac{\alpha_{ij}}{\alpha_{ij}^0}, \qquad \frac{\sum_j \alpha_{ij} A_{ij}}{\sum_j \alpha_{ij}^0 A_{ij}} \tag{S1}$$

each averaged with equal weight over samples, heads and queries at every lead, on the base times at which all six leads are available, so that every lead is measured on the same inputs; a pull ratio below one means that the distance term moves the weight towards the upwind position. On the 96-station network (six folds, eight seeds) $\gamma$ falls from 1.61 at 1 h to 0.15 at 48 h, the relative size from 1.55 to 0.12 and the divergence from 0.78 to 0.017 nats, while the pull ratio rises from 0.49 to 0.94; the relative size and the divergence decrease monotonically with lead in 47 of the 48 runs. The content scale is 1.31 at 1 h and 1.21 at 3 h and rises to 1.88 at 48 h; it is monotone in 15 of the 48 runs. On the 724-station network (six folds, five seeds) $\gamma$ falls from 3.22 to 0.06, the relative size from 3.09 to 0.06 and the divergence from 1.76 to 0.0030 nats, and the pull ratio rises from 0.26 to 0.98; the relative size and the divergence decrease monotonically in all 30 runs, and the content scale (1.16–1.56) is monotone in one. The decay of $\gamma(\tau)$ is therefore not offset by the content scale, which grows at long leads: on both networks the relative weight of the distance term decays with lead.

**Synthetic control.** The data and the intervention are defined in the Methods. In lead order (1, 3, 6, 12, 24 and 48 h), the self-input coefficient $a(\tau)$ is 0.9, 0.6, 0.4, 0.5, 0.6 and 0.7, and the independent Gaussian noise has standard deviation 0.2, 0.3, 0.4, 0.4, 0.5 and 0.6. Each data set has 4,500 training, 780 calibration and 550 evaluation base samples, each at all six leads. The synthetic model uses three variables, zero tendencies and a single zero-valued static channel, without temporal-phase features; its optimization settings match the main experiments. Inputs are standardized using the 1 h training inputs. The first variable at each of the five selected neighbours is increased by 0.1 standardized units, and the change in the target prediction is divided by 0.1 to report the effect per unit perturbation. Each lead and seed has 4,400 interventions, and every target has five non-self stations in its intervention set. With the dependence present, the rank correlation across leads between the true effect and the model effect is 0.94–1.00 in the three seeds, that between the true effect and the relative size 0.89–1.00, that between the true effect and $\gamma$ 0.14–0.54, and that between the true effect and the divergence −0.26. In the control data set ($\omega \equiv 0$) the three seeds learn different coefficients: $\gamma$ stays below 0.004 at every lead in one seed, reaches 3.40 at 48 h in the second and 24.3 in the third, where the divergence reaches 15.8 nats; in all three the mean model effect is below 0.0001 in magnitude at every lead, and the mean absolute effect is below 0.00044.

**Joint scaling of the coefficient and the content logits (Fig. 5f).** On the 48 stored 96-station models, $\gamma(\tau)$ was multiplied at inference by 0.25, 0.5, 1, 2 or 4 and the content logits by 0.5, 1 or 2, and the quantities of equation (S1) and the energy-score effect of the distance term — the energy score with the distance term removed minus that with it — were recomputed on the same rows. Multiplying both by the same factor leaves the relative size unchanged but changes the divergence and the energy-score effect: at 6 h, halving both, leaving both unchanged and doubling both give divergences of 0.056, 0.207 and 0.609 nats and energy-score effects of 0.0015, 0.0083 and 0.0203. At a fixed multiplier of $\gamma$ the divergence changes little with the logit multiplier (0.056 and 0.057 nats at 6 h for $\gamma$ halved and logits halved or unchanged), whereas the energy-score effect grows with it (0.0015 and 0.0055). Neither the coefficient nor its ratio to the content scale therefore determines the effect of the distance term on the forecast by itself.

## Supplementary Note 4. Effect of the covariance floor

The Cholesky diagonal is bounded below at 0.05 in standardized units for numerical stability. At 1 h the two pressures are strongly over-dispersed before and after recalibration (PIT variance 0.21 of the uniform value before and 0.31–0.32 after recalibration; Fig. 3b of the main text). A model retrained with the floor lowered to 0.005 was compared with the full model on the same seeds (0–2) and six multi-year folds. With the original floor the 1 h CRPS is 0.225 hPa for station pressure and 0.234 hPa for sea-level pressure, against 0.220 and 0.230 hPa for EMOS; with the lowered floor it is 0.200 and 0.205 hPa, below that of EMOS. Lowering the floor improves the joint negative log-likelihood by 0.49–0.96 nats per station and raises $\mathbb{E}[z^2]$ of the standardized pressure difference from 0.09–0.14 to 0.34–0.54, closer to its ideal of one. The energy score, however, is worse at every lead, by 3.5% at 1 h and 0.8–2.9% at 3–48 h, and its variogram score is worse at every lead as well. Under the lowered floor the learned correlations attain a better log-likelihood than the fixed empirical correlation at 3–48 h, whereas under the original floor the fixed correlation is better at every lead ('Joint uncertainty and recalibration' in the main text). The original floor is retained because the energy score is the primary metric; the floor sets a trade-off between the joint likelihood, the pressure marginals and the kernel scores.

## Supplementary Note 5. Evaluation of the dependence structure

The contribution of the learned dependence structure is measured post hoc from the stored predictions of the multi-year models, with no retraining. Three predictive distributions share the predictive mean and the marginal variances of each forecast and differ only in the correlation matrix: the learned covariance (learned correlations), its diagonal (conditional independence), and a covariance whose correlation matrix is a fixed per-lead empirical correlation of the standardized residuals $z_v = (y_v - \mu_v)/\sqrt{\Sigma_{vv}}$ of the training split (fixed correlation). Each is scored before and after the lead-wise scale $s(\tau)$; the main text reports the issued distributions. The joint negative log-likelihood is computed in closed form, the energy score with the common random numbers of the main evaluation (two batches of 1,000 draws), and the variogram score of order 0.5 over the ten variable pairs, with unit weights, from the first batch.

Derived quantities are evaluated in physical units. The pressure difference $\mathrm{ps} - \mathrm{pa}$ is linear in the variables, so its predictive distribution is Gaussian; its CRPS is computed in closed form and its calibration assessed using $\mathbb{E}[z^2] = \mathbb{E}[(a^\top(y-\mu))^2/(a^\top\Sigma a)]$, with $a$ its coefficient vector and ideal value 1. The discomfort index $0.81\,T + 0.01\,H\,(0.99\,T - 14.3) + 46.3$ and wind chill $13.12 + 0.6215\,T - 11.37\,U^{0.16} + 0.3965\,T\,U^{0.16}$, with $T$ in °C, $H$ in % and $U$ the wind speed in km h$^{-1}$, are evaluated from the 1,000 draws of the first batch, and their calibration is read from the variance of the probability integral transform of the observation relative to the uniform value $1/12$ (ties counted half). Wind chill is defined only for $T \leq 10$ °C and $U > 3$ mph (about 4.83 km h$^{-1}$), so it is scored only on rows whose observation lies in that domain, from the draws inside the domain, and rows with fewer than 50 such draws under any of the three dependence structures, before or after recalibration, are excluded, so that all are compared on the same rows. The compound hot–dry event is temperature above the 90th percentile and humidity below the 10th percentile of the training split, with the percentiles computed over all stations and target times of that split; its probability is the fraction of draws in the event, scored by the Brier score. A Brier score computed from a Monte Carlo event fraction and a variogram score computed from a Monte Carlo mean are biased by the sampling variance of the draws. We remove the Brier-score bias by subtracting $\hat{p}(1-\hat{p})/(n_s - 1)$ from each squared error. For the variogram score, we replace the squared Monte Carlo mean with the product of the means from the two independent batches (seed 0 of each multi-year fold). After these corrections, the relative differences between the three dependence structures, and between the two covariance floors, change by less than 0.061 percentage points and retain their sign at every lead.

## Supplementary Note 6. Related approaches

This note gives in full the comparison summarized under 'Related approaches' in the Methods.

**Multi-lead weather prediction.** Predicting several leads with one model requires an input that identifies the target time of each forecast. Supplying the lead to the network as an extra input encoding is widely used (Sønderby et al., 2020; Espeholt et al., 2022). Where in the network the lead enters also affects accuracy (Nguyen et al., 2024), and attention weights have been analysed to interpret what a spatio-temporal model relies on (Liang et al., 2024).

Recent work predicts several leads with one shared model rather than building a separate model per lead (Bashir et al., 2026; Liu & Rahnemoonfar, 2026). Both, however, address deterministic prediction, and neither treats the advection information available on a station network or the joint predictive distribution of several variables.

Station observations have also been used to supply local information. Miralles et al. (2025) proposed a graph neural network for nowcasting that combines station observations with radar and numerical weather prediction (NWP), and a subsequent multimodal ablation found that station observations improve local and rain-onset diagnostics, local skill being a distinct target from gridded-field skill (Miralles et al., 2026). The MetNet family began from radar and satellite and extended the range from 8 h to 24 h while adding temperature, dew point and wind as predictands (Sønderby et al., 2020; Espeholt et al., 2022; Andrychowicz et al., 2023). Aardvark Weather and AIFS-DOP reduce the dependence on reanalysis further and forecast out to ten days from observations (Allen et al., 2025; Pinnington et al., 2026). A probabilistic extension of such a system has also been proposed (Almeida et al., 2026). These systems nonetheless use assimilated fields or radar alongside the observations, or issue per-variable marginal distributions or deterministic forecasts only.

Pinnington et al. (2026) note that squared-error training can over-smooth the predicted field and leave a probabilistic loss to future work.

One study unifies stations worldwide in a single deep model and infers propagation direction from the learned correlations (Wu et al., 2023); its axis of unification is the station, and its output a point forecast (a reference run on the data of this study is reported in Supplementary Note 2). On a sparse network, a mixture-of-experts graph transformer routes over weather regimes to predict three variables out to 3–12 h (Li et al., 2027), where the routing variable is the regime rather than the lead.

Studies that use station observations alone to issue a joint predictive distribution over several surface variables across multiple leads, and that analyse how the usefulness of spatial information varies with lead, remain limited.

**Physically informed graph attention and neural advection operators.** Graph attention learns from data how much each neighbour contributes to the forecast at a target station. Prior information such as distance, position or spatial covariance has been added to the attention score to impose spatial structure. The Spatially-Informed Transformer learns a geostatistical covariance that varies with inter-station distance and incorporates it into the attention score (Calleo, 2025); it accounts for spatial relations but addresses a single variable. iTransformer instead treats each variable as a token and learns the relations among variables by attention (Liu, Y., et al., 2024); we use it, with the same joint output head attached, as a principal baseline for cross-variable relations (Supplementary Note 9).

Graph-based methods that set the direction of the physical link between stations from the wind have also been developed. Graph attention learns the relative importance of connected nodes from data (Veličković et al., 2018), and graph networks in weather use that structure to exchange information between stations or grid cells (Lam et al., 2023). Introducing wind direction lets the observation at an upwind station weigh more in the forecast downwind. Existing wind-informed graph models, however, mostly address single-variable nowcasting, and rarely adjust the strength with which upwind observations are used as a function of lead.

Neural advection and transport operators predict the future state by moving the current field along the flow. de Bézenac et al. (2019) learned advection and diffusion with a network to warp the input field, and Shi et al. (2017) used location-dependent recurrent connections to evolve the spatio-temporal state. ClimODE integrates a continuous-time advection equation with a learned velocity field from the base time and emits a per-variable Gaussian mean and variance (Verma et al., 2024). These approaches resemble optical-flow nowcasting, which estimates the displacement of precipitation areas in radar imagery (Ayzel et al., 2019). They can represent the motion of weather systems directly, but generally advance one forecast state from step to step, either feeding each step's forecast into the next or integrating it forward in time; error from an earlier step can then accumulate at later leads, and how the usefulness of upwind observations varies with lead is hard to establish separately.

Rather than advancing the base-time state step by step, CLARA computes the upwind position required at each lead directly. From the target position $p_i$, the base-time wind $w_i$ and the lead $\tau$ it defines the upwind position, the point reached from $p_i$ by moving against $w_i$ for a time $\tau$ along the sphere, and incorporates the distance between that position and each neighbouring station into the attention score. No forecast is fed forward: the forecast at each lead is computed from the observations at the same base time.

CLARA also does not apply the same strength of upwind information at every lead. Defining that strength as a learnable function of lead, $\gamma(\tau)$, lets the contribution of spatial advection information be adjusted lead by lead, and the learned $\gamma(\tau)$ curve describes how the coefficient of the upwind-distance bias changes with lead. Because attention weighs the curve against the content-path scale, the curve is read together with that scale, which is measured directly (Lemma 1 in the Methods; Supplementary Note 3).

CLARA combines this treatment of spatial information with multivariate probabilistic prediction in one model. It produces the mean and covariance of five variables at each station, forming a joint predictive distribution that includes the dependence among variables, evaluates it by the energy score, and recalibrates the magnitude of its uncertainty per lead.

Work on the identifiability of attention addresses the relation between weights and outputs (Brunner et al., 2020) or the invariance of the query–key bilinear form, not the learned coefficient of an additive bias. Linear position bias fixes the slope (Press et al., 2022), and the same work reports that a learnable slope did not improve on the fixed one, without stating a cause. Attention logits carry no unobserved error term, so the scale normalization of discrete choice does not carry over to attention: the content scale can be computed directly rather than normalized away.

**Multivariate probabilistic forecasting and uncertainty calibration.**

Predictive distributions in probabilistic weather forecasting are constructed in three broad ways: by generating ensemble members, by correcting the bias and uncertainty of forecasts already produced, and by having the model output the parameters of the distribution — a mean and a covariance — directly. The three differ in how they express predictive uncertainty and dependence among variables.

Generating members covers physical ensembles as well as models that learn exchangeable members with a proper scoring rule (Lang et al., 2024; Pacchiardi et al., 2024). Lakatos (2026) generates members with a graph neural network instead of estimating a covariance, calibrating the multivariate dependence with a loss combining the energy and variogram scores. Learning the marginals first and restoring the spatial joint structure afterwards (Alet et al., 2025), and sampling with diffusion models, belong to the same category. These methods represent complex distributions flexibly, but computing a probability requires generating many members.

Where a neural post-processor takes an ensemble as input, how the members enter matters. Höhlein et al. (2024) feed the full set of members to a permutation-invariant network rather than summary statistics such as mean and variance, preserving the information the ensemble carries, but their target is per-station post-processing of a single variable. Both Lakatos (2026) and Höhlein et al. (2024) require an NWP or ensemble forecast as input and so differ from a system operated from station observations alone.

Representative post hoc methods are EMOS and neural post-processing (Gneiting et al., 2005; Rasp & Lerch, 2018), which mostly calibrate the marginal distribution of each variable. Where dependence across variables or stations is needed, calibrated marginals are commonly recombined by ensemble copula coupling (Schefzik et al., 2013). A bivariate EMOS instead fits a truncated bivariate normal with a full covariance to wind speed and temperature from ensemble statistics (Baran & Möller, 2017). Conformal calibration can complement CLARA by calibrating joint prediction regions (Asch et al., 2026). With the fitted predictive mean and covariance held fixed, calibration of the squared Mahalanobis score can define ellipsoidal regions with finite-sample marginal coverage guarantees under exchangeability of calibration and test scores (Angelopoulos & Bates, 2023). This assumption is not established for the seasonal evaluations here. Region coverage alone does not specify the predictive dependence structure or the distributions of derived quantities.

Post-processing models are often fitted separately per variable, station and lead, typically with diurnal and seasonal predictors formed by cyclic functions of hour and day of year (Ylinen et al., 2020). To reduce per-lead fitting, Wessel et al. (2024) represented the calibration coefficients as functions varying smoothly with lead — again for single-variable post-processing of an NWP ensemble. Assouline et al. (2026) applied a lead-aware bias correction while downscaling machine-learning forecasts to kilometre scale. Defining calibration coefficients or fusion weights as functions of lead is thus used across several settings, but, to our knowledge, has not been carried over to issuing a multivariate joint predictive distribution from observations alone.

With a Gaussian parameterization, the network directly outputs the predictive mean and covariance. In multivariate distributional regression it is established practice to parameterize the covariance through a Cholesky factor so that it has a valid structure; Muschinski et al. (2024) applied this to joint probabilistic forecasts of temperature across several leads. In general multivariate time-series forecasting, a recurrent network emits the mean and a low-rank-plus-diagonal covariance of a Gaussian copula across series and draws multi-step forecasts by sampling (Salinas et al., 2019), and a network can emit the Cholesky factor of a conditional copula correlation across the forecast horizons of one series, fitted after separately learned quantile functions (Wen & Torkkola, 2019); both address non-meteorological series. ClimODE emits a variance for each variable but no covariance between variables (Verma et al., 2024). What is then available without generating members depends on what is emitted: marginal variances, a latent copula dependence, or, as in Muschinski et al. (2024) and CLARA, a full predictive covariance.

Standard benchmarks such as EUPPBench support the development of multivariate probabilistic forecasting and post-processing (Demaeyer et al., 2023), but they presuppose a post-processing problem in which an NWP ensemble is supplied, and therefore do not address the setting of this study, in which the predictive distribution is produced from station observations without NWP or reanalysis.

CLARA follows the third route while extending it so that the parameters of the distribution reflect neighbouring stations and the lead. It produces the mean and covariance of five variables at each station, computing the covariance from an attention representation that has integrated information from neighbours, and then applies a lead-wise scale $s(\tau)$ estimated on a separate calibration split to recalibrate the magnitude of the covariance while preserving the correlation structure. The estimator is the multivariate, lead-wise form of the moment matching used in σ-scaling of regression uncertainty (Levi et al., 2022; Laves et al., 2020) and in covariance-inflation estimation for ensemble data assimilation (Li et al., 2009). The conditions under which this estimator is consistent, and its statistical properties, are stated in Proposition 1 of the Methods; the full proof, assumptions and further results are given in Supplementary Note 11.

Extended Data Table 1 compares prior systems with CLARA by input data, output and resolution, lead range, probabilistic output, lead-time handling and reliance on NWP or reanalysis. Among the systems compared we find none that issues a joint predictive distribution over several variables from station observations alone while adjusting both the strength of an upwind-position bias and the magnitude of the covariance by lead.

CLARA is evaluated on six multi-year rolling-origin folds and two 2026 held-out seasons of two nationwide networks, with statistical and learned baselines. The advected objective-analysis reference covers both nationwide networks, an adapted Corrformer reference uses the six multi-year folds of the 96-station network, and regional replication covers ten networks.

## Supplementary Note 7. Definition of ablation and decomposition variants

Every ablation variant uses the architecture, data, folds, seeds and training settings of the full model and differs from it in one respect; the decompositions below are defined relative to the same full model. The full model uses both the lead-conditioned advection bias $\gamma(\tau)A_{ij}$ and the temporal-phase input. No-advection removes the distance term from the attention logit and keeps the spatial attention. No-phase removes the six temporal-phase features from the station token. Constant-γ holds $\gamma$ at one learned constant, $\gamma = \mathrm{softplus}(\beta_\gamma)$ with $\beta_\gamma$ a learnable scalar. On both networks, zero-wind sets the wind vector used for the upwind location to zero in training and evaluation, so that the distance term uses the plain distance between stations; it was built from the same caches with only the wind components changed. The contribution of a component is the fold-mean energy score of the variant minus that of the full model, so a positive value means that the component improves performance.

Two decompositions locate where the lead information is used. Gateonly holds the lead channel of the token embedding at zero, so that explicit lead information enters only through $\gamma(\tau)$. Constonly additionally holds $\gamma$ at a learned constant, so that no explicit lead input remains in the token embedding or the coupling. Both retain the target-hour phase and the lead-proportional upwind displacement, so the lead remains partly recoverable: the target-hour phase recovers it modulo 24 h, and wherever the wind is non-zero the upwind displacement also separates 24 h from 48 h. Holding the lead channel at zero encodes every sample as if its lead were 6 h ($\tau_{\mathrm{norm}} = 0$).

The per-lead specialist trains a separate model on the samples of one lead, with the same number of optimization steps as the full model of the same fold and seed, so that each specialist passes over its data about six times as often. The floor variant lowers the Cholesky diagonal floor from 0.05 to 0.005 and is otherwise identical to the full model.

## Supplementary Note 8. Network architecture and training details

The station token of CLARA takes as input the current observation $y_i$, the tendency over the past 60 min $\Delta y_i$, station elevation, the normalized lead time and the temporal phase information. These inputs pass through a linear layer of latent dimension $d_{\mathrm{model}} = 64$ with a GELU (Gaussian error linear unit) activation and become one station token. Spatial combination between station tokens uses a single multi-head self-attention block with four heads (16 dimensions per head); the attention logit receives the salience offset $b_j$, computed for each head by a linear map of the key station's token, and the lead-conditioned advection bias. The attention output is updated through a sigmoid-gated residual connection, layer normalization and a feed-forward network. The final station token feeds the mean head and the Cholesky head. To ensure a valid covariance matrix, the diagonal elements of the Cholesky factor are bounded below by adding 0.05 to the softplus output. The negative log-likelihood (NLL) loss never forms the inverse of the covariance matrix explicitly: the quadratic form is computed as $\| L_i(t,\tau)^{-1}\left(y_{i,t+\tau} - \mu_i(t,\tau)\right) \|^2$ by a triangular solve, and the log-determinant as $\log\det\Sigma_i(t,\tau) = 2\sum_v \log L_{i,vv}(t,\tau)$. The positive lower bound on the diagonal keeps the log term finite. The implementation omits the constant $V/2\log 2\pi$ of equation (7), which changes neither the optimum nor the quoted differences between conditions.

The multilayer perceptron (MLP) that produces the advection coupling $\gamma(\tau)$ has a 1–16–1 structure — one input unit, 16 hidden units and one output — with a GELU activation on the hidden layer. To represent a wide range of times stably, the lead $\tau$ is converted to minutes and then log-transformed, centred and scaled by fixed constants as $\tau_{\mathrm{norm}} = (\ln\tau - \ln 360)/0.9$. The transformed lead enters both the station token and the multilayer perceptron that produces $\gamma(\tau)$.

In the domestic experiments the model uses station elevation as its only static attribute and has 27,979 trainable parameters; the regional experiments use elevation, latitude and longitude and a three-variable output head (Supplementary Note 12), with the same architecture otherwise.

Continuous inputs are standardized with the mean and the standard deviation computed from the base-time observations of each fold's training split. Targets are standardized with the same statistics, and the energy score is computed in that standardized space. One batch holds 18 sample groups (one base time and lead each). The model was trained for 40 epochs with the Adam optimizer at a learning rate fixed at $1.5 \times 10^{-3}$, with no learning-rate schedule; weight decay, dropout and early stopping were not used, and no hyperparameter was tuned on evaluation data. Training and inference times are given in Supplementary Note 14.

**Training and forecast issuance procedure of CLARA**

- **Input**: for training, the targets $y_i(t+\tau)$; for recalibration, held-out calibration data (step 7); for training and issuance, base-time observations $y_i(t)$ and 60-min tendencies $\Delta y_i(t)$ (five variables), station elevation, station coordinates, the set of leads (1–48 h), temporal-phase features, and wind observations for the upwind distance $d_{ij}^{\mathrm{up}}$.
- **Output**: trained parameters and, for every station $i$ and lead $\tau$, the issued predictive law $\mathcal{N}(\mu_i(t,\tau),\ s(\tau)^2\Sigma_i(t,\tau))$ (main-text equation (9)).
- (1) Tokenization: map the inputs of each station through the linear layer of width $d_{\mathrm{model}} = 64$ with GELU into a station token.
- (2) Advection-biased attention: score each neighbour by the content score $\ell_{ij}^{\mathrm{content}}(\tau)$ plus the per-key offset $b_j$ minus the upwind-distance term $\gamma(\tau)\ d_{ij}^{\mathrm{up}}/100$ (main-text equation (2) with the offset $b_j$ added), with the coupling $\gamma(\tau) = \mathrm{softplus}(\mathrm{MLP}(\tau_{\mathrm{norm}}))$ (equation (3); $\tau_{\mathrm{norm}}$ as defined above); update the tokens through the sigmoid-gated residual, layer normalization and the feed-forward layer.
- (3) Output heads: the mean head produces the increment $\delta_i(t,\tau)$; the Cholesky head produces the 15 lower-triangular elements, with softplus + 0.05 on the diagonal.
- (4) Assembly: $\mu_i = y_{i,t} + \delta_i$ (equation (5)) and $\Sigma_i = L_i L_i^{\mathrm{T}}$ (equation (6)).
- (5) Loss: evaluate the NLL of equation (7) through the triangular quadratic form $\| L_i^{-1}(y_{i,t+\tau} - \mu_i) \|^2$ and average over the batch.
- (6) Optimization: repeat (1)–(5) with Adam (learning rate fixed at $1.5 \times 10^{-3}$, batches of 18 sample groups) for 40 epochs.
- (7) Recalibration: on the held-out calibration split, estimate the per-lead scale $s(\tau) = \sqrt{\widehat{\mathbb{E}}_{\mathrm{calib}}[d^2]/V}$ (equation (8); $d^2, V$: Supplementary Note 13).
- (8) Issuance: for each evaluation base time, one forward pass per lead issues $\mathcal{N}(\mu_i,\ s(\tau)^2\Sigma_i)$ (equation (9)).

## Supplementary Note 9. Baseline implementations

Six external baselines and one learned alternative were used: persistence, which carries the current observation forward; climatology; EMOS (ensemble model output statistics) fitted per lead; iTransformer, which attends along the variable axis, without the temporal-phase input and, as the learned alternative, with it; graphAttn, which applies self-attention across stations; and a physical baseline that advects an objective analysis built by interpolating observations onto a grid. All statistical baselines are fitted on the training split alone, using the rows of that split that satisfy the scoring criteria, and every probabilistic baseline is recalibrated with its own lead-wise scale $s(\tau)$ fitted on the calibration split, as CLARA is. The probabilistic baselines are scored on the same base times, stations, leads and targets as CLARA, in the same standardized space; the physical baseline is deterministic and is treated separately below.

The persistence baseline takes the base-time observation as the predictive mean and the per-lead residual covariance of the training split as the predictive covariance. The climatology baseline uses the mean and covariance of the training-split samples for each combination of lead and target hour, and falls back to per-lead statistics when a combination has fewer than 40 samples. EMOS is adapted here to an observation-only setting in which no ensemble is available, so the mean comes from a regression and the spread from its residuals rather than from ensemble members (Gneiting et al., 2005). It regresses the target for each variable and lead by least squares on seven predictors: a constant, persistence, the tendency, and the sine and cosine of the target-hour and day-of-year phases. Its predictive covariance is the per-lead empirical residual covariance. The phases are among its inputs, as they are for CLARA, whereas the default iTransformer and graphAttn baselines do not receive them.

graphAttn uses the architecture of CLARA with both the advection bias and the temporal-phase input removed, following the graph spatio-temporal attention family (Liang et al., 2024). It can therefore learn relations between stations from the data, but makes no explicit use of the upwind location or of the diurnal and seasonal information of the target time. Because it takes no phase input its embedding layer is 384 parameters smaller: 27,595 against CLARA's 27,979. iTransformer attaches the same joint output head to a variable-axis attention backbone and is trained with the same negative log-likelihood, seeds (eight on 96 ASOS, five on 724 AWS) and optimization settings as CLARA; it has 18,352 trainable parameters without and 18,640 with the phase input.

Against the no-phase learned baselines on 96 ASOS, lead-mean improvements are 6.2% over graphAttn and 16.5% over iTransformer; their inputs also differ (Supplementary Table 2). Because iTransformer and graphAttn without phase input differ from CLARA in input information as well as architecture, they are reported in Supplementary Table 2, and iTransformer + phase provides a learned comparison with matched temporal-phase input. The original learned baselines were not parameter-count matched. To obtain a parameter count similar to CLARA, we increased the width of iTransformer + phase from 48 to 60 (27,616 parameters versus 27,979) and retrained it using the same training, recalibration and evaluation protocol. Across the six multi-year folds, the lead-mean energy score of CLARA's issued forecasts was 4.7% lower on 96 ASOS and 6.2% lower on 724 AWS; before recalibration the reductions were 5.1% and 6.6%. At every lead the issued fold-mean score favoured CLARA; the direction held in all six folds at 1–24 h and five of six at 48 h on both networks. Figure 2a and Extended Data Fig. 1a use the original width-48 model; Fig. 2b and Extended Data Fig. 1b also include the retrained width-60 model.

The physical baseline advects an observation-based surface objective analysis at 500 m resolution over the country, using the observed station wind at the base time. At each target station the upwind location is computed with the spherical formula used by CLARA, and the analysis value in the nearest grid cell, if that cell lies within 500 m, is taken as the forecast in a semi-Lagrangian manner. Where the upwind location falls outside the analysis grid the forecast is treated as missing, so the fraction of scorable rows is 0.81 at 1 h and 0.43 at 48 h on the 96-station network, and 0.79 and 0.31 on the 724-station network. This baseline produces a deterministic forecast, and it is compared with CLARA on exactly the rows on which it issues a forecast (Supplementary Table 1). The comparison covers both nationwide networks (Supplementary Fig. 2).

The operational NWP reference was evaluated on the GHCNh networks of two regions. Over the US Midwest, the extended cycles of the High-Resolution Rapid Refresh (HRRR; 00, 06, 12 and 18 UTC) were matched to the base times (age zero), and 2 m temperature, 2 m relative humidity and the 10 m wind components at forecast hours 1–48 h were extracted at the nearest grid point of each station. Over the Korean Peninsula, the ECMWF ensemble (control plus 50 perturbed members, 51 in total) and the high-resolution deterministic run (HRES) were obtained from the TIGGE (THORPEX Interactive Grand Global Ensemble) archive at the 00 and 12 UTC cycles, and the forecast hours available on its 6-h grid (6, 12, 24 and 48 h) were matched to the base times in the same way. The reduced Gaussian grid was interpolated bilinearly to the stations, and relative humidity was computed from temperature and dew point by the Magnus approximation.

The variables compared are temperature, wind speed and humidity, as in the regional experiments, on the GHCNh stations within 3 km of a station at which the NWP product was extracted. The deterministic comparison uses the per-variable RMSE of HRRR, of the ensemble mean and of HRES against the seed-mean prediction of CLARA. The probabilistic comparison uses the fair CRPS (continuous ranked probability score, with the finite-ensemble bias removed) of the members against the closed-form CRPS of the marginals of the issued CLARA distribution, recalibrated on the calibration split as in the other experiments. Because the CRPS of a deterministic forecast equals its mean absolute error (MAE), the probabilistic comparison against HRRR is CLARA's CRPS against the HRRR MAE. Both NWP systems may assimilate observations from the same stations that are available at initialization and differ from CLARA in their inputs, so they are read as operational reference points rather than as competitors under matched conditions (Supplementary Tables 6 and 7, Extended Data Fig. 3).

## Supplementary Note 10. Rolling-origin validation protocol and calibration split

The evaluation comprises eight seasonal folds on each domestic network: six multi-year folds with evaluation seasons in summer (JJA) and winter (DJF) of 2023, 2024 and 2025, and two held-out folds with evaluation periods December 2025–February 2026 and June–August 2026. Each fold's training window spans the 23–29 months before its evaluation season (per-fold boundaries and sample counts in Supplementary Table 8). The last 15% of the training window, split at a calendar boundary, forms the calibration split, which is excluded from training and from the fitting of every baseline and of the standardization statistics and is used only to fit $s(\tau)$. A base-time–lead pair belongs to a split only if both its base time and its target time lie inside that split, so no training or calibration target falls in a later split. The recalibration fallback to a global scale for leads with fewer than 50 calibration rows was never used for a scored lead: every lead of every run trained on all leads had at least 9,900 calibration rows.

The two 2026 seasons were held out: they were not used for model development or for choosing any setting. For each of the two 2026 folds, models were trained from scratch on that fold's preceding observations using the fixed architecture, training settings and calibration procedure, without transferring weights from the multi-year folds.

For winter 2026, training spans 1 January 2024–17 August 2025, calibration 18 August–30 November 2025, and evaluation 1 December 2025–28 February 2026. The two domestic networks each have 3,576 training groups and 624 calibration groups; evaluation retains 548 groups on 96 ASOS and 540 on 724 AWS. Of the 92 scheduled base times, 92 and 91, respectively, retain at least one evaluable lead (Supplementary Table 8).

The configuration had been chosen in earlier experiments on 2024–2025 data from AWS stations in central Korea, which overlap the 2024 and 2025 multi-year folds in time. The 2026 folds were not used to select CLARA's configuration or the pre-specified primary comparisons; Corrformer and the regional learned models were evaluated in additional comparisons. The parameter-count comparison retrained iTransformer + phase at increased width under the same training, recalibration and evaluation protocol; these additional comparisons were not part of the pre-specified primary comparisons. Because the multi-year folds share training windows and there are only six of them, year-to-year results are reported through effect sizes and the number of folds with a given sign rather than by a formal significance test. The regional evaluation uses the same eight folds with the dates read in UTC (Supplementary Note 12).

## Supplementary Note 11. Assumptions and proofs of Proposition 1 and Corollary 1

This note states the assumptions under which the recalibration estimator of the Methods is consistent and gives the full proofs of Proposition 1 and Corollary 1. It also derives the pseudo-true scale under general misspecification (part (ii)), establishes the probability-integral-transform result (part (iv)), provides the trace-based counterexample (part (v)) and reports their numerical verification.

**Assumptions.**

Four assumptions are used below. (A1) The predictive mean is correct. (A2) The conditional law of the observation is Gaussian with the stated covariance. (A3) At a given lead the calibration and evaluation periods share the same scale $c(\tau)$, so that $d^2$ has the same distribution in both. (A4) At a fixed lead the sampled days — each a block of the four base hours 00/06/12/18, with days sampled every four days — are independent or α-mixing with $\sum_k \alpha_{\text{mix}}(k)^{\xi/(2+\xi)} < \infty$ for some $\xi > 0$, and each day contributes a bounded number of samples. Of these, (A1) and (A2) enter the proofs of Proposition 1 and Corollary 1, (A3) transfers a calibration-split estimate to the evaluation split in the Methods, and (A4) supports the law of large numbers used in part (ii) of the proof of Proposition 1 and in equation (S12) of Corollary 1. Assumption (A3) is testable. In the main text ('Joint uncertainty and recalibration') the scale fitted on the calibration split, the last 15% of the training window, over-corrects at 12–48 h in the evaluation season, which is consistent with (A3) failing at long leads when the two periods differ in season. The algebraic identities and estimator properties below were also verified numerically. Correlation preservation and the softmax identity hold to machine precision ($\sim 10^{-16}$). The multivariate PIT is uniform within sampling error over 20 replications of $2 \times 10^5$ draws (Kolmogorov–Smirnov statistic 0.0020 on average against a 5% critical value of 0.0030). The per-variable scales of Corollary 1 recover the true diagonal factors to within Monte Carlo error. The energy-score estimator's bias is within one Monte Carlo standard error of zero, with variance scaling as $1/n_s$ (ratio 3.98 against the predicted 4 when $n_s$ is quadrupled).

**Proof of Proposition 1 and Corollary 1.** Let $\Sigma = LL^\top$ be the model covariance with non-singular Cholesky factor $L$ (its diagonal is bounded below by 0.05 in the implementation), and suppose $y \mid x \sim \mathcal{N}(\mu, c(\tau)^2\Sigma)$ while the model predicts $\mathcal{N}(\mu, \Sigma)$, with $c(\tau) > 0$. Write $d^2 = (y-\mu)^\top\Sigma^{-1}(y-\mu)$, with $V$ the dimension of $y$. Parts (i)–(v) below establish, in order, the distribution of $d^2$ and the consistency of $s(\tau)$, the exactness of the recalibration and the preservation of the correlation matrix, the PIT property, and the fact that $\mathbb{E}[d^2] = V$ is necessary but not sufficient; the diagonal extension is established by Corollary 1 at the end of this note.

(i) Distribution of $d^2$. Since, by (A1)–(A2), $y - \mu = c\, L\varepsilon$ with $\varepsilon \sim \mathcal{N}(0, I_V)$,

$$d^2 = (c\,L\varepsilon)^\top (LL^\top)^{-1}(c\,L\varepsilon) = c^2\,\varepsilon^\top L^\top (L^\top)^{-1}L^{-1}L\,\varepsilon = c^2\,\varepsilon^\top\varepsilon \;\sim\; c^2\,\chi^2_V, \tag{S2}$$

so $\mathbb{E}[d^2] = c^2V$ and $\mathrm{Var}[d^2] = 2c^4V$.

(ii) Consistency. On the calibration split the $n$ per-lead calibration samples $(d^2)_m$ all have mean $c^2V$ and, being scaled $\chi^2_V$ variables, finite moments of every order. They are dependent within a sampled day, whose stations share one synoptic situation and whose four base hours are 6 h apart, but by (A4) only weakly dependent across days. Grouping the samples into the $B$ day blocks with block sums $T_g = \sum_{m\in g}(d^2)_m$, each over at most $K$ samples by (A4), the covariance inequality for α-mixing sequences gives $\left|\mathrm{Cov}(T_g, T_{g'})\right| \le 8\,\alpha_{\mathrm{mix}}(|g-g'|)^{\xi/(2+\xi)}\|T_g\|_{2+\xi}\|T_{g'}\|_{2+\xi}$ for $g \ne g'$, while $\mathrm{Var}(T_g) \le \|T_g\|^2_{2+\xi}$; with $\|T_g\|_{2+\xi} \le K\|d^2\|_{2+\xi}$ this yields

$$\mathrm{Var}(s^2) = \frac{1}{n^2V^2}\sum_{g,g'}\mathrm{Cov}\,(T_g, T_{g'}) \le \frac{8K^2\|d^2\|^2_{2+\xi}}{n^2V^2}\,B\Big(1 + 2\sum_{k\ge1}\alpha_{\mathrm{mix}}\,(k)^{\xi/(2+\xi)}\Big) = O(1/B) \tag{S3}$$

since $n \ge B$ and the series converges by (A4). Chebyshev's inequality then gives, as $B \to \infty$,

$$s(\tau)^2 = \frac{1}{nV}\sum_{m=1}^{n}(d^2)_m \;\xrightarrow{p}\; c(\tau)^2, \qquad \text{hence} \quad s(\tau) \;\xrightarrow{p}\; c(\tau) \tag{S4}$$

by the continuous mapping theorem. Under independent sampling the central limit reference rate is $\sqrt{n}\,(s^2 - c^2) \Rightarrow \mathcal{N}(0,\, 2c^4/V)$; with dependent samples the variance also depends on within-day and between-day covariances, so this independent-sampling rate is not used quantitatively. Without (A1)–(A2), the same argument with uniformly bounded $(2+\xi)$-moments of $d^2$ gives $s(\tau)^2 - c^2_{\star n} \xrightarrow{p} 0$ for the pseudo-true value

$$c^2_{\star n} = \frac{1}{nV}\sum_{m=1}^{n}\Big[\mathrm{tr}\big(\Sigma_m^{-1}\Sigma_{\star,m}\big) + b_m^\top\Sigma_m^{-1}b_m\Big] \tag{S5}$$

where $b_m$ is the mean error and $\Sigma_{\star,m}$ the true covariance of calibration sample $m$; under the scalar form of (A1)–(A2) this pseudo-true value equals $c(\tau)^2$.

(iii) Exact recalibration and correlation preservation. With $\Sigma' = s^2\Sigma$ the corrected statistic is

$$d'^2 = (y-\mu)^\top(s^2\Sigma)^{-1}(y-\mu) = d^2/s^2, \qquad s = c \;\Rightarrow\; d'^2 \sim \chi^2_V, \quad \mathbb{E}[d'^2] = V. \tag{S6}$$

For any scalar $s > 0$,

$$\mathrm{corr}(\Sigma')_{uv} = \frac{s^2\Sigma_{uv}}{\sqrt{s^2\Sigma_{uu}}\sqrt{s^2\Sigma_{vv}}} = \mathrm{corr}(\Sigma)_{uv}, \tag{S7}$$

so the correlation matrix is unchanged.

(iv) PIT exactness. Under a correctly specified model ($c = 1$), $d^2 \sim \chi^2_V$ whose CDF $G_V$ is continuous; by the probability integral transform $G_V(d^2) \sim \mathrm{U}(0,1)$.

(v) Necessity, not sufficiency. For a correct mean and true covariance $\Sigma_\star$, any predicted $\tilde\Sigma$ gives

$$\mathbb{E}[d^2] = \mathbb{E}\,\mathrm{tr}\big(\tilde\Sigma^{-1}(y-\mu)(y-\mu)^\top\big) = \mathrm{tr}\big(\tilde\Sigma^{-1}\Sigma_\star\big), \tag{S8}$$

so the calibration constraint is $\mathrm{tr}(\tilde{\Sigma}^{-1}\Sigma_\star) = V$, which does not determine $\tilde{\Sigma}$. A counterexample with $V = 2$ and $\Sigma_\star = I_2$:

$$\tilde{\Sigma} = \mathrm{diag}(2,\ 2/3): \qquad \mathrm{tr}\big(\tilde{\Sigma}^{-1}\big) = 1/2 + 3/2 = 2 = V, \qquad \tilde{\Sigma} \neq I_2. \tag{S9}$$

Corollary 1 (per-variable diagonal extension). For $\kappa_v(\tau) > 0$, let

$$y \mid x \sim \mathcal{N}\big(\mu,\ R^{1/2}\Sigma R^{1/2}\big), \qquad R = \mathrm{diag}\,\kappa_v(\tau) \tag{S10}$$

Then the standardized residual $z_v = (y_v - \mu_v)/\sqrt{\Sigma_{vv}}$ satisfies

$$\mathbb{E}[z_v^2] = \frac{\big(R^{1/2}\Sigma R^{1/2}\big)_{vv}}{\Sigma_{vv}} = \kappa_v, \tag{S11}$$

so, by the law of large numbers of (A4),

$$s_v(\tau)^2 = \widehat{\mathbb{E}}_{\mathrm{calib}}[z_v^2] \ \xrightarrow{p}\ \kappa_v(\tau) \tag{S12}$$

(the scale identity uses second moments), and at the population value $S = R^{1/2}$ the corrected covariance $S\Sigma S$ equals the true covariance exactly. If the misspecification were of this diagonal form, the corrected covariance would also be calibrated jointly: with $d_S^2 = (y-\mu)^\top (S\Sigma S)^{-1}(y-\mu)$, equation (S8) gives

$$\Sigma_\star = R^{1/2}\Sigma R^{1/2}, \quad S = R^{1/2} \quad \Rightarrow \quad \mathbb{E}[d_S^2] = \mathrm{tr}((S\Sigma S)^{-1}\Sigma_\star) = \mathrm{tr}(I_V) = V \tag{S13}$$

## Supplementary Note 12. Observations, quality control and preprocessing

Training and evaluation target the five-variable joint distribution at stations at which all five variables are observed. The inputs of a sample are all stations at which the five variables (ta, ws, hm, pa, ps) are valid at the base time, and a sample is formed only if there are at least five such stations. The loss and every score use only the input stations at which all five variables are also valid at the target time, so the choice of inputs does not depend on the target. The inputs are the current observation $y_i$ at each station (the main text's $y_{j,t}$, keeping only the station subscript), the 60 min tendency $\Delta y_i$, station elevation, the normalized lead time and the temporal phase. The standardization statistics (mean and standard deviation) are computed from the base-time observations of each fold's training split alone. The four base hours {00, 06, 12, 18} Korea Standard Time (KST) within a sampled day are treated as one dependent block (assumption (A4) in Supplementary Note 11).

Three station-observation datasets, each with a different analytical purpose, were used (Extended Data Table 7). Records from the 96 Automated Synoptic Observing System (ASOS) stations provide the main evaluation, the calibration and dependence analyses and the ablations. Records from 724 automatic weather station (AWS) sites nationwide test whether the results are reproduced on a denser network of a different type. GHCNh records from ten regions test whether the results are reproduced in other climates and networks.

The nominal number of AWS sites is about 7.5 times the ASOS count, but the number of sites reporting all five variables is 4.6–6.3 times the ASOS count in the fold means of the scored stations per sample.

The comparison between the two networks therefore changes not only the density but also the type of network, and it evaluates whether the accuracy and the operating characteristics of the model are reproduced on a different network, not the isolated effect of density.

In the domestic records, values outside physically plausible ranges were set to missing: temperature −35 to 45 °C, wind speed 0–60 m $s^{-1}$, relative humidity 0.1–100%, station and sea-level pressure 800–1,080 hPa and wind direction 0–360°. Missing records were handled as follows. Where the observation 60 min before the base time was absent at a station, its tendency was set to zero. Where the wind direction was missing, both wind components were set to zero; the upwind location then coincides with the station, so the advection bias reduces to an ordinary distance bias. Sensitivity to missing neighbours — the stations that supply context — was evaluated separately ('Replication across observing networks' in the main text, Extended Data Fig. 2).

Regional replication ('Replication across regional observing networks' in the main text) was evaluated on the NOAA Global Historical Climatology Network hourly (GHCNh) archive, the successor of the Integrated Surface Database, retrieved as station-year files on 27 September 2026. The ten regions on six continents span Köppen classes A, B, C and D; each region is a fixed latitude–longitude box that does not follow national borders. Stations were selected before any model was trained and without using 2026 data: a station qualified if the GHCNh inventory records at least one observation in every month of 2021–2025 and if, in each year from 2021 to 2025, temperature, dew point, wind speed and wind direction were all valid within 20 min before at least half of the base hours (00, 06, 12 and 18 UTC); one representative was taken per cell of an 8 × 8 grid over the box; farthest-point sampling reduced the set to 50 where more stations qualified, and where fewer than 50 cells had a representative, further candidates were added in a fixed random order. Central Brazil (36 stations), East Africa (13) and southeastern Australia (15) have fewer qualifying stations.

Quality flags follow the GHCNh documentation (lower-case alphabetic flags and the numeric flags that mark failed checks are rejected), and when several reports have the same timestamp the synoptic report is kept over the aviation reports. Inputs are the latest valid observation within 20 min before the base time, and targets are the nearest report within ±20 min of the target time; a target whose actual observation time falls outside its validation window is treated as missing.

The eight folds are those of the domestic experiments with the same dates read in UTC, with base times every four days at 00, 06, 12 and 18 UTC. The joint distribution comprises three variables (temperature, wind speed and relative humidity computed from temperature and dew point), because station and sea-level pressure are reported at base hours by fewer than half of the qualifying stations in several regions. Three seeds were trained per condition, and the architecture, optimization settings and covariance floor are those of the domestic experiments; the static attributes are elevation, latitude and longitude. The diurnal phase uses local solar time referenced to the mean longitude of the selected stations of each region. In forming the upwind location, the wind displacement is applied along the great circle on a sphere of radius 6,371 km, and station separations are great-circle (haversine) distances; the same formula is used in every region.

An additional regional comparison retrained iTransformer + phase with width 64 and three output variables (27,335 parameters, versus 27,136 for regional CLARA), with three seeds per fold. Across the six multi-year folds, CLARA had a 2.8–5.4% lower lead-mean issued energy score in each of the ten regions and a lower score in 57 of 60 region–lead means. The exceptions were EU at 48 h (−0.20% improvement), KR at 24 h (−0.03%) and KR at 48 h (−1.10%); at KR 48 h, only one of six folds favoured CLARA. Before recalibration, all 60 region–lead means favoured CLARA. These comparisons use the existing fold-mean-of-seed-means aggregation, followed by equal weighting of six leads; percentage reductions are computed from the aggregated scores, not by averaging percentages.

In both 2026 seasons, CLARA has a lower issued energy score than persistence in every region–lead combination. Against the width-64 learned comparator, CLARA has a lower score in 54 of the 60 region–lead means. Its lead-mean reduction is 3.0–5.0% in nine regions; in central Brazil the reduction is −1.2%, favouring the comparator. The six region–lead means favouring the comparator are BR at 1, 3, 24 and 48 h, EU at 48 h and US at 48 h.

## Supplementary Note 13. Joint calibration diagnostics

Joint calibration is summarized by the mean of the Mahalanobis statistic $d^2 = (y-\mu)^\top \Sigma^{-1}(y-\mu)$, whose ideal value is $V = 5$, and the shape of the distribution is diagnosed by the multivariate probability integral transform (PIT) — the $\chi^2_5$ CDF applied to $d^2$, uniform when the Gaussian joint law is correctly specified; uniformity alone does not establish joint calibration. Per-variable calibration is assessed using the PIT. For the diagonal extension below, define the standardized residual $z_v = (y_v - \mu_v)/\sqrt{\Sigma_{vv}}$. The issued distributions use the per-lead scalar $s(\tau) = \sqrt{\widehat{\mathbb{E}}_{\mathrm{calib}}[d^2]/V}$, and a single global scalar $s$ is reported for comparison (Extended Data Table 3). The per-variable diagonal of Corollary 1, $S = \operatorname{diag} s_v(\tau)$ with $s_v(\tau)^2 = \widehat{\mathbb{E}}_{\mathrm{calib}}[z_v^2]$, is derived but not evaluated here. All three preserve the correlation matrix exactly: the scalars because a common factor cancels, and the diagonal because correlation is invariant to per-variable rescaling.

## Supplementary Note 14. Computational cost

Computational cost was measured with a stored full model (96 ASOS, summer 2025 fold, seed 0; 27,979 parameters) on one workstation with an Intel Core i9-12900 processor and an NVIDIA GeForce RTX 3080 GPU, using PyTorch 2.5.1, with no other workload running (Supplementary Table 9). Inference is one forward pass for one base time and one lead, covering all input stations of that sample (median 96), timed as the median over 200 evaluation samples after 20 warm-up passes; the time to assemble the inputs, including the distance and upwind matrices, is reported separately. Training is one complete 40-epoch run from new weights on the training split of the same fold (4,508 sample groups in batches of 18).

A forward pass takes 0.43 ms on one CPU thread, 0.39 ms on eight threads and 1.24 ms on the GPU, so the forecasts for all 552 sample groups of the evaluation season take 0.24 s on one thread. Training takes 2.1 min on one thread, 1.6 min on eight threads and 1.7 min on the GPU. Peak host memory is 655 MB and 666 MB in the one- and eight-thread CPU runs and 1,281 MB in the GPU run, whose peak GPU allocation is 79 MB. Power was not measured.

The experiments were run with PyTorch 1.13.1 and CUDA 11.7, mostly on an NVIDIA RTX A5000 GPU. All randomness (weight initialization, batch order and Monte Carlo draws) was generated from fixed seeds with deterministic algorithms enforced, so rerunning the same command in the same environment reproduces the same weights and results. A GPU is not required for training or prediction.

## Supplementary Tables

Supplementary Table 1. Advected observation-based objective analysis (OA) versus CLARA on the 96 ASOS and 724 AWS stations: coverage (fraction of scored rows on which the OA issues a forecast) and per-variable RMSE in physical units on the same rows (fold mean over the six rolling-origin folds; RMSE from the fold-mean squared error).

| Network / lead | Coverage | Temperature (°C) OA / CLARA | Wind speed (m/s) OA / CLARA | Humidity (%) OA / CLARA | Station pressure (hPa) OA / CLARA | Sea-level pressure (hPa) OA / CLARA |
|---|---|---|---|---|---|---|
| **96 ASOS** | | | | | | |
| 1 h | 0.81 | 1.61 / 0.69 | 1.19 / 0.73 | 7.69 / 4.16 | 15.17 / 0.36 | 0.67 / 0.37 |
| 3 h | 0.83 | 2.95 / 1.26 | 1.44 / 0.91 | 12.69 / 7.20 | 23.59 / 0.70 | 1.42 / 0.74 |
| 6 h | 0.82 | 4.70 / 1.82 | 1.70 / 1.06 | 19.68 / 9.73 | 28.09 / 1.16 | 1.97 / 1.21 |
| 12 h | 0.77 | 6.30 / 2.42 | 1.86 / 1.20 | 26.26 / 12.14 | 32.17 / 1.94 | 3.02 / 2.00 |
| 24 h | 0.65 | 4.18 / 2.62 | 1.74 / 1.27 | 17.53 / 12.86 | 35.55 / 3.34 | 4.45 / 3.41 |
| 48 h | 0.43 | 4.86 / 3.26 | 1.84 / 1.39 | 18.15 / 13.57 | 37.25 / 4.63 | 5.95 / 4.73 |
| **724 AWS** | | | | | | |
| 1 h | 0.79 | 1.90 / 0.76 | 1.38 / 0.78 | 9.08 / 4.40 | 20.97 / 0.41 | 0.78 / 0.41 |
| 3 h | 0.77 | 3.16 / 1.30 | 1.60 / 0.99 | 13.96 / 7.38 | 26.23 / 0.71 | 1.48 / 0.74 |
| 6 h | 0.73 | 4.88 / 1.87 | 1.80 / 1.14 | 20.23 / 9.99 | 28.53 / 1.13 | 1.94 / 1.18 |
| 12 h | 0.65 | 6.42 / 2.47 | 1.99 / 1.26 | 25.82 / 12.70 | 30.31 / 1.89 | 2.89 / 1.96 |
| 24 h | 0.51 | 4.07 / 2.68 | 1.83 / 1.27 | 18.41 / 13.17 | 31.54 / 3.32 | 4.39 / 3.41 |
| 48 h | 0.31 | 4.67 / 3.39 | 1.92 / 1.42 | 18.78 / 13.95 | 31.76 / 4.67 | 5.88 / 4.80 |

*Note. RMSE ↓, lower is better. The OA is a deterministic forecast issued only where the upwind point lies inside the 500 m analysis grid; CLARA's predictive mean (seed mean of the squared error) is scored on exactly those rows. The station-pressure error of the OA reflects the elevation gap between grid cell and station.*

Supplementary Table 2. Learned baselines without temporal-phase input: energy score (fold mean) of the issued distribution by lead time, with the improvement of CLARA and the number of folds in which CLARA performs better in parentheses; 96 ASOS (eight seeds) and 724 AWS (five seeds), six rolling-origin folds.

| Lead | 96 ASOS iTransformer | 96 ASOS graphAttn | 724 AWS iTransformer | 724 AWS graphAttn |
|---|---|---|---|---|
| 1 h | 0.295 (+5.4%, 6/6) | 0.283 (+1.3%, 5/6) | 0.283 (+7.2%, 6/6) | 0.269 (+2.3%, 6/6) |
| 3 h | 0.483 (+18.6%, 6/6) | 0.418 (+6.0%, 6/6) | 0.473 (+20.3%, 6/6) | 0.403 (+6.5%, 6/6) |
| 6 h | 0.673 (+26.9%, 6/6) | 0.568 (+13.3%, 6/6) | 0.662 (+28.0%, 6/6) | 0.545 (+12.6%, 6/6) |
| 12 h | 0.845 (+28.4%, 6/6) | 0.674 (+10.2%, 6/6) | 0.839 (+29.5%, 6/6) | 0.651 (+9.2%, 6/6) |
| 24 h | 0.745 (+7.0%, 6/6) | 0.710 (+2.4%, 6/6) | 0.730 (+7.9%, 6/6) | 0.689 (+2.4%, 6/6) |
| 48 h | 0.887 (+7.9%, 6/6) | 0.844 (+3.2%, 5/6) | 0.868 (+8.6%, 6/6) | 0.819 (+3.1%, 6/6) |

*Note. ES ↓, lower is better. iTransformer (variable-axis attention) and graphAttn (station self-attention on the CLARA backbone without the advection-routing bias) receive no temporal-phase input, so these comparisons differ from CLARA in input information as well as architecture.*

Supplementary Table 3. Reference comparison with Corrformer on the 96 ASOS multi-year folds: energy score of the issued distribution (fold mean of seed means, three seeds) of Corrformer, CLARA and persistence on the same rows, the improvement of CLARA, the number of folds in which CLARA performs better and in which Corrformer performs

better than persistence, and the root-mean-square error of the predictive mean in standardized units on the same rows (fold mean of seed-mean squared errors).

| Lead | Energy score Corrformer / CLARA / persistence | Improvement of CLARA over Corrformer (%) | Folds CLARA better (ES) | Folds Corrformer better than persistence (ES) | RMSE, standardized Corrformer / CLARA / persistence | Folds CLARA better (RMSE) |
|---|---|---|---|---|---|---|
| 1 h | 0.429 / 0.278 / 0.330 | 35.3 | 6/6 | 0/6 | 0.310 / 0.219 / 0.249 | 6/6 |
| 3 h | 0.501 / 0.393 / 0.554 | 21.6 | 6/6 | 4/6 | 0.362 / 0.302 / 0.401 | 6/6 |
| 6 h | 0.569 / 0.493 / 0.804 | 13.4 | 6/6 | 6/6 | 0.408 / 0.372 / 0.576 | 6/6 |
| 12 h | 0.657 / 0.609 / 1.058 | 7.2 | 6/6 | 6/6 | 0.468 / 0.452 / 0.745 | 5/6 |
| 24 h | 0.768 / 0.697 / 0.842 | 9.3 | 6/6 | 6/6 | 0.536 / 0.512 / 0.592 | 6/6 |
| 48 h | 0.904 / 0.817 / 1.005 | 9.6 | 6/6 | 6/6 | 0.632 / 0.592 / 0.705 | 6/6 |

*Note. ES ↓, lower is better; $I = 100\,(S_{Corrformer} - S_{CLARA})/S_{Corrformer}$. Corrformer (width 128, 23,949,071 parameters) issues point forecasts; its predictive distribution combines its point forecast with the per-lead training-split residual covariance recalibrated with s(τ), as for persistence and EMOS (Supplementary Note 2).*

Supplementary Table 4. Comparison under a common recalibration: lead-mean improvement of CLARA over each baseline and the number of fold–lead cells in which CLARA performs better, for the issued distributions (every model recalibrated with its own lead-wise scale s(τ)) and before recalibration (raw), on the multi-year folds.

| Network | Baseline | Lead-mean improvement (%) issued / raw | Fold–lead cells CLARA better issued / raw |
|---|---|---|---|
| 96 ASOS | iTransformer + phase | 4.9 / 5.3 | 35/36 / 35/36 |
| 96 ASOS | Persistence | 28.8 / 28.4 | 36/36 / 36/36 |
| 96 ASOS | Climatology | 64.7 / 64.8 | 36/36 / 36/36 |
| 96 ASOS | EMOS | 11.0 / 10.3 | 36/36 / 36/36 |
| 96 ASOS | iTransformer (no phase) | 16.5 / 16.9 | 36/36 / 36/36 |
| 96 ASOS | graphAttn (no phase) | 6.2 / 6.3 | 34/36 / 34/36 |
| 724 AWS | iTransformer + phase | 7.0 / 7.2 | 35/36 / 36/36 |
| 724 AWS | Persistence | 28.4 / 27.9 | 36/36 / 36/36 |
| 724 AWS | Climatology | 65.8 / 65.8 | 36/36 / 36/36 |
| 724 AWS | EMOS | 12.2 / 11.5 | 36/36 / 36/36 |
| 724 AWS | iTransformer (no phase) | 17.7 / 18.0 | 36/36 / 36/36 |
| 724 AWS | graphAttn (no phase) | 6.0 / 6.1 | 36/36 / 35/36 |

*Note. $I = 100\,(S_{ref} - S_{CLARA})/S_{ref}$ of the lead-mean energy scores, positive when CLARA performs better. The ranking is the same with and without recalibration.*

Supplementary Table 5. Replication of the multi-year comparison on the 724-station AWS network: energy score (fold mean) of the issued distribution by lead time, with the improvement of CLARA in parentheses; six rolling-origin folds, five seeds for CLARA.

| Lead | CLARA | iTransformer + phase | Persistence | Climatology | EMOS | Advected OA |
|---|---|---|---|---|---|---|
| 1 h | 0.263 | 0.274 (+4.1%) | 0.310 (+15.3%) | 1.536 (+82.9%) | 0.285 (+7.9%) | 0.844 (+70.2%) |
| 3 h | 0.377 | 0.414 (+9.1%) | 0.532 (+29.1%) | 1.540 (+75.5%) | 0.451 (+16.4%) | 1.176 (+69.6%) |
| 6 h | 0.476 | 0.530 (+10.2%) | 0.777 (+38.7%) | 1.535 (+69.0%) | 0.593 (+19.7%) | 1.472 (+69.4%) |
| 12 h | 0.591 | 0.653 (+9.4%) | 1.021 (+42.1%) | 1.536 (+61.5%) | 0.709 (+16.6%) | 1.827 (+69.5%) |
| 24 h | 0.673 | 0.708 (+5.0%) | 0.822 (+18.1%) | 1.546 (+56.5%) | 0.733 (+8.3%) | 1.544 (+60.5%) |
| 48 h | 0.794 | 0.834 (+4.8%) | 0.973 (+18.4%) | 1.575 (+49.6%) | 0.844 (+5.9%) | 1.656 (+56.7%) |

*Note. ES ↓, lower is better; improvement I = 100 ($S_{ref}$ − $S_{CLARA}$)/$S_{ref}$, positive when CLARA performs better. CLARA's fold-mean energy score is lower than that of every listed model at all six evaluated leads and lower in 143 of the 144 fold–lead cells against iTransformer + phase and the three statistical baselines. In the two 2026 held-out seasons, CLARA's seed-mean energy score is lower than that of the five pre-specified baselines (persistence, climatology, EMOS, iTransformer and graphAttn) in all 60 season–lead–baseline cells. iTransformer + phase receives the same temporal-phase input as CLARA. Learned models without phase input are reported in Supplementary Table 2. The advected OA is a deterministic physical baseline scored only on the rows whose upwind point lies inside the analysis grid (coverage 0.31–0.79 by lead); its column compares CLARA on those same rows.*

Supplementary Table 6. Operational NWP reference over the US Midwest: HRRR (3 km, hourly assimilation, extended cycles matched at the base time) against CLARA on shared stations and times.

| Lead | Rows | Temperature (°C) RMSE HRRR / CLARA | Wind speed (m/s) RMSE HRRR / CLARA | Humidity (%) RMSE HRRR / CLARA | Temperature (°C) MAE HRRR / CRPS CLARA | Wind speed (m/s) MAE HRRR / CRPS CLARA | Humidity (%) MAE HRRR / CRPS CLARA |
|---|---|---|---|---|---|---|---|
| 1 h | 2825 | 1.10 / 0.85 | 1.39 / 1.20 | 7.29 / 4.35 | 0.81 / 0.45 | 1.07 / 0.65 | 5.64 / 2.30 |
| 3 h | 2821 | 1.35 / 1.43 | 1.46 / 1.51 | 8.42 / 6.72 | 1.00 / 0.76 | 1.15 / 0.82 | 6.46 / 3.57 |
| 6 h | 2803 | 1.62 / 2.07 | 1.56 / 1.81 | 9.28 / 8.84 | 1.17 / 1.10 | 1.18 / 0.98 | 7.08 / 4.73 |
| 12 h | 2797 | 1.84 / 2.72 | 1.63 / 2.11 | 10.64 / 10.41 | 1.33 / 1.47 | 1.23 / 1.15 | 8.11 / 5.68 |
| 24 h | 2794 | 1.94 / 3.60 | 1.70 / 2.34 | 11.94 / 11.36 | 1.43 / 1.95 | 1.29 / 1.29 | 9.20 / 6.26 |
| 48 h | 2769 | 2.29 / 4.78 | 1.92 / 2.33 | 13.72 / 13.19 | 1.74 / 2.64 | 1.45 / 1.31 | 10.46 / 7.29 |

*Note. Lower is better. HRRR is deterministic, so its CRPS equals its MAE. Shared stations are the selected GHCNh stations within 3 km of the stations at which HRRR was extracted; only the winter 2024 and summer 2025 folds have HRRR extractions. HRRR is not post-processed to stations and assimilates observations hourly; CLARA is trained and recalibrated on the stations.*

Supplementary Table 7. Operational NWP reference over the Korean Peninsula: the raw ECMWF ensemble (TIGGE) and its high-resolution run (HRES) against CLARA on shared stations and times (six multi-year folds).

| Lead | Rows | Temperature (°C) CRPS ensemble / CLARA | Wind speed (m/s) CRPS ensemble / CLARA | Humidity (%) CRPS ensemble / CLARA | Temperature (°C) RMSE ens. mean / HRES / CLARA | Wind speed (m/s) RMSE ens. mean / HRES / CLARA | Humidity (%) RMSE ens. mean / HRES / CLARA |
|---|---|---|---|---|---|---|---|
| 6 h | 8117 | 1.28 / 0.97 | 1.12 / 0.74 | 7.04 / 5.18 | 1.90 / 1.89 / 1.83 | 1.78 / 1.78 / 1.40 | 10.76 / 10.78 / 9.78 |
| 12 h | 8160 | 1.36 / 1.01 | 1.06 / 0.78 | 7.49 / 5.85 | 1.98 / 1.98 / 1.82 | 1.70 / 1.71 / 1.51 | 11.28 / 11.32 / 10.56 |
| 24 h | 8110 | 1.34 / 1.38 | 1.10 / 0.90 | 7.65 / 6.88 | 1.98 / 1.98 / 2.47 | 1.79 / 1.80 / 1.72 | 11.69 / 11.83 / 12.27 |
| 48 h | 7947 | 1.39 / 1.71 | 1.08 / 0.96 | 7.90 / 7.82 | 2.10 / 2.11 / 3.04 | 1.83 / 1.88 / 1.86 | 12.34 / 12.62 / 13.89 |

*Note. Lower is better. The ensemble CRPS uses all 51 members (control and 50 perturbed); the ensemble is not post-processed. Base times 00 and 12 UTC; leads on the 6 h TIGGE grid. Shared stations are the selected GHCNh stations within 3 km of the stations at which TIGGE was extracted.*

Supplementary Table 8. Rolling-origin folds: training, calibration and evaluation periods (each from its start date to the start date of the next, end exclusive), sample groups per split (base time × lead), evaluation base times, scored stations per sample and distinct evaluation stations.

| Fold | Training | Calibration | Evaluation | Groups train / calib / eval | Base times (eval) | Scored stations per sample (mean) | Distinct evaluation stations |
|---|---|---|---|---|---|---|---|
| **96 ASOS** | | | | | | | |
| winter2023 | 2021-01-01 – 2022-08-19 | 2022-08-19 – 2022-12-01 | 2022-12-01 – 2023-03-01 | 3576 / 624 / 548 | 92 | 95.4 | 96 |
| summer2023 | 2021-01-01 – 2023-01-20 | 2023-01-20 – 2023-06-01 | 2023-06-01 – 2023-09-01 | 4501 / 781 / 552 | 92 | 95.3 | 96 |
| winter2024 | 2022-01-01 – 2023-08-19 | 2023-08-19 – 2023-12-01 | 2023-12-01 – 2024-03-01 | 3576 / 624 / 552 | 92 | 95.3 | 96 |
| summer2024 | 2022-01-01 – 2024-01-21 | 2024-01-21 – 2024-06-01 | 2024-06-01 – 2024-09-01 | 4508 / 788 / 552 | 92 | 95.7 | 96 |
| winter2025 | 2023-01-01 – 2024-08-18 | 2024-08-18 – 2024-12-01 | 2024-12-01 – 2025-03-01 | 3576 / 624 / 548 | 92 | 94.6 | 96 |
| summer2025 | 2023-01-01 – 2025-01-20 | 2025-01-20 – 2025-06-01 | 2025-06-01 – 2025-09-01 | 4508 / 788 / 552 | 92 | 95.3 | 96 |
| winter2026 | 2024-01-01 – 2025-08-18 | 2025-08-18 – 2025-12-01 | 2025-12-01 – 2026-03-01 | 3576 / 624 / 548 | 92 | 95.4 | 96 |
| summer2026 | 2024-01-01 – 2026-01-20 | 2026-01-20 – 2026-06-01 | 2026-06-01 – 2026-09-01 | 4508 / 788 / 552 | 92 | 95.7 | 96 |
| **724 AWS** | | | | | | | |
| winter2023 | 2021-01-01 – 2022-08-19 | 2022-08-19 – 2022-12-01 | 2022-12-01 – 2023-03-01 | 3576 / 624 / 548 | 92 | 435.6 | 455 |
| summer2023 | 2021-01-01 – 2023-01-20 | 2023-01-20 – 2023-06-01 | 2023-06-01 – 2023-09-01 | 4501 / 781 / 552 | 92 | 449.2 | 468 |
| winter2024 | 2022-01-01 – 2023-08-19 | 2023-08-19 – 2023-12-01 | 2023-12-01 – 2024-03-01 | 3576 / 624 / 552 | 92 | 504.3 | 527 |
| summer2024 | 2022-01-01 – 2024-01-21 | 2024-01-21 – 2024-06-01 | 2024-06-01 – 2024-09-01 | 4508 / 788 / 552 | 92 | 511.2 | 526 |
| winter2025 | 2023-01-01 – 2024-08-18 | 2024-08-18 – 2024-12-01 | 2024-12-01 – 2025-03-01 | 3576 / 624 / 548 | 92 | 560.7 | 578 |
| summer2025 | 2023-01-01 – 2025-01-20 | 2025-01-20 – 2025-06-01 | 2025-06-01 – 2025-09-01 | 4508 / 788 / 552 | 92 | 563.7 | 578 |
| winter2026 | 2024-01-01 – 2025-08-18 | 2025-08-18 – 2025-12-01 | 2025-12-01 – 2026-03-01 | 3576 / 624 / 540 | 91 | 601.1 | 614 |
| summer2026 | 2024-01-01 – 2026-01-20 | 2026-01-20 – 2026-06-01 | 2026-06-01 – 2026-09-01 | 4508 / 780 / 552 | 92 | 601.5 | 615 |

*Note. winter2026 and summer2026 are the two held-out folds; the others are the six multi-year folds. Base times every four days at 00, 06, 12 and 18 KST; a group is one base time and lead. The calibration split is the last 15% of each training window and is excluded from training.*

Supplementary Table 9. Computational cost of CLARA (27,979 parameters, 96 ASOS summer 2025 fold) on one workstation, CPU and GPU, PyTorch 2.5.1: inference for one base time and one lead covering all input stations of the sample (median 96), the forward passes for the 552 sample groups of the evaluation season, one complete 40-epoch training run and peak memory.

| Device | Threads | Forward pass (ms) median / 90th pct. | With input assembly (ms) | Evaluation season forward (s) | 40-epoch training (min) | Peak host memory (MB) | Peak GPU memory (MB) |
|---|---|---|---|---|---|---|---|
| CPU (Intel Core i9-12900) | 1 | 0.43 / 0.46 | 0.69 | 0.24 | 2.1 | 655 | — |
| CPU (Intel Core i9-12900) | 8 | 0.39 / 0.47 | 0.66 | 0.22 | 1.6 | 666 | — |
| GPU (NVIDIA GeForce RTX 3080) | — | 1.24 / 5.50 | 1.70 | 0.69 | 1.7 | 1,281 | 79 |

*Note. Inference is the median over 200 evaluation samples after 20 warm-up passes; the input assembly includes the distance and upwind matrices. Training starts from new weights on the training split (4,508 sample groups, batches of 18). No other workload was running.*

## Supplementary Figures

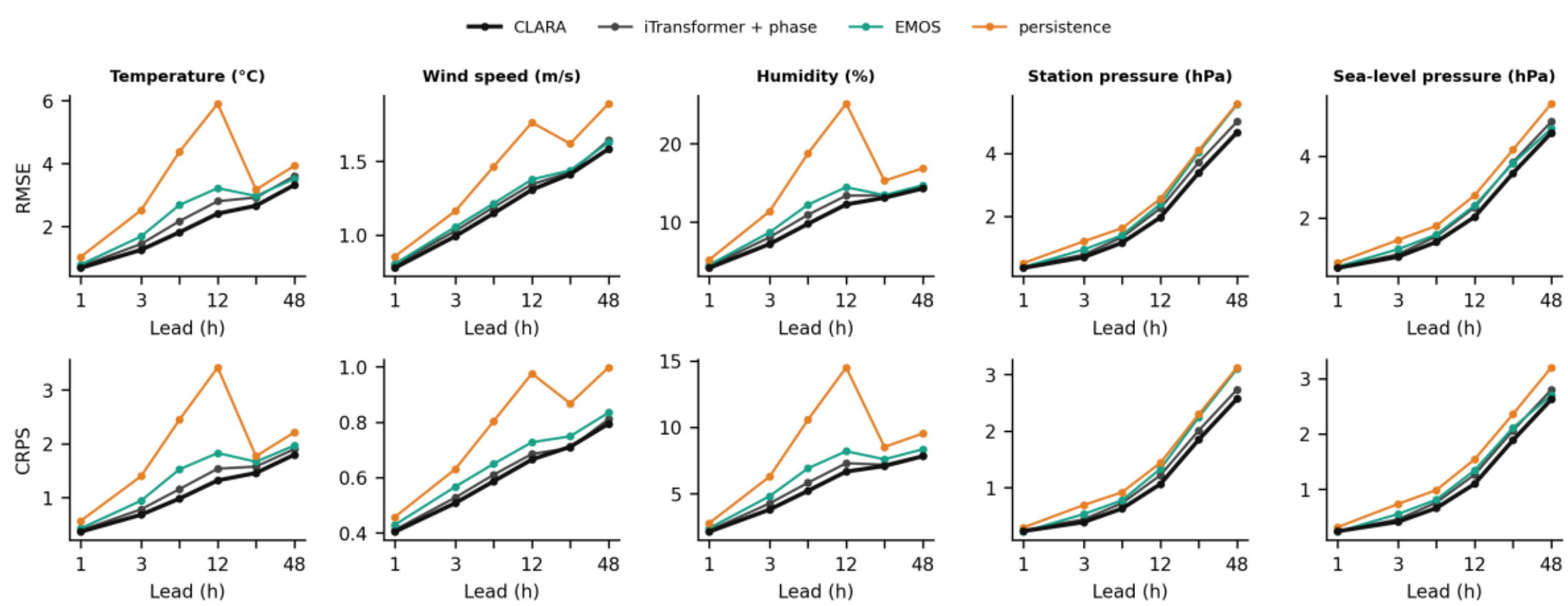

*Supplementary Figure 1. Per-variable RMSE (top) and CRPS of the issued distribution (bottom) in physical units by lead for CLARA, iTransformer + phase, EMOS and persistence (96 ASOS, six multi-year folds; RMSE from the fold-mean squared error).*

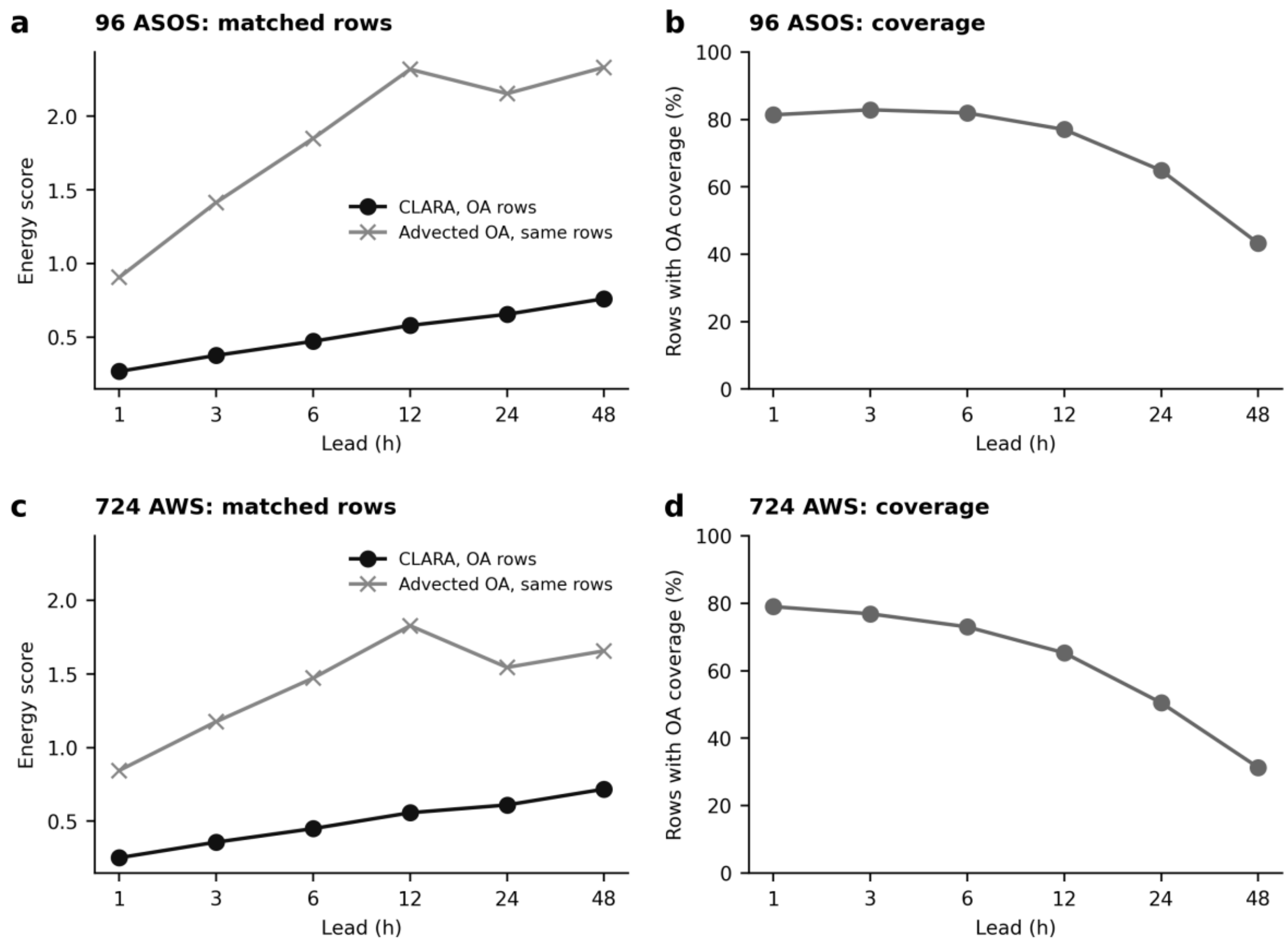

*Supplementary Figure 2. Matched-row comparisons with advected objective analysis (OA) and its coverage on the two station networks (six multi-year folds).* ***a, c****, Energy scores of issued CLARA forecasts and deterministic OA forecasts, both evaluated on the same OA-covered rows, for 96 ASOS and 724 AWS, respectively.* ***b, d****, The fraction of evaluation rows with valid OA forecasts by lead for the corresponding networks. Scores and coverage are fold means; CLARA scores are first averaged over seeds within each fold.*

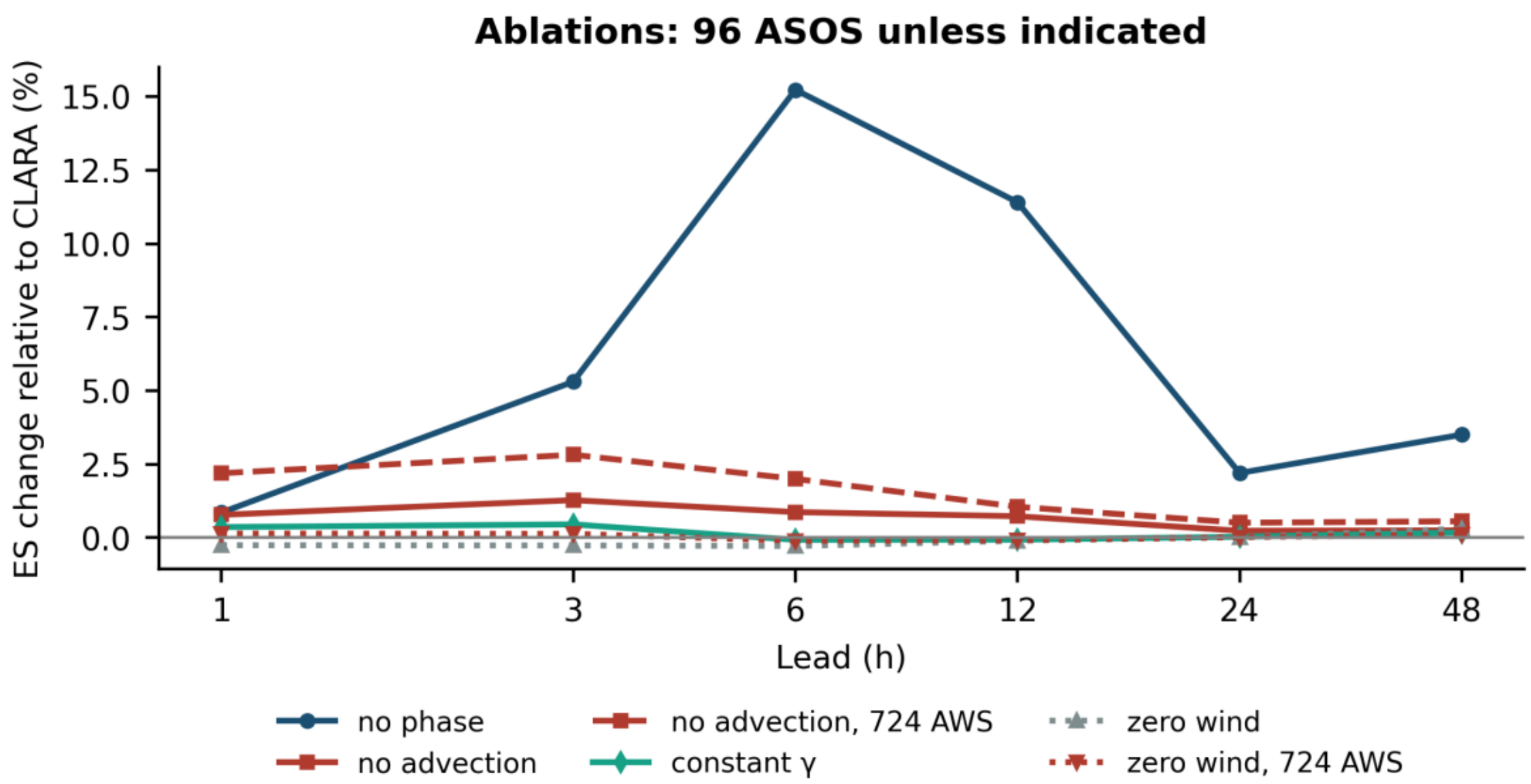

*Supplementary Figure 3. Retrained ablations of CLARA. Energy-score change relative to CLARA after removing temporal phase, removing advection, making γ constant or setting the wind in the distance bias to zero. Results are for 96 ASOS, except for the separately labelled 724 AWS no-advection and zero-wind series. Positive values indicate a higher energy score for the ablated model. All series use the issued distributions; numerical results are in Extended Data Table 5.*

## Supplementary References

*References cited in this Supplementary Information; formats follow the main text.*